\documentclass[a4paper,oneside,10pt]{article}%
\usepackage{amsmath}
\usepackage{amsfonts}
\usepackage{amssymb}
\usepackage{graphicx}
\usepackage[square,numbers,sort&compress]{natbib}%
\providecommand{\U}[1]{\protect\rule{.1in}{.1in}}
\newtheorem{theorem}{Theorem}[section]

\newtheorem{definition}[theorem]{Definition}
\newtheorem{assumption}[theorem]{Assumption}
\newtheorem{example}[theorem]{Example}

\newtheorem{lemma}[theorem]{Lemma}

\newtheorem{proposition}[theorem]{Proposition}
\newtheorem{remark}[theorem]{Remark}

\numberwithin{equation}{section}

\begin{document}

\title{Recursive utility optimization with concave coefficients}
\author{Shaolin Ji \thanks{Institute for Financial Studies, Shandong University, Jinan
250100, China and Institute of Mathematics, Shandong University, Jinan 250100,
China, Email: jsl@sdu.edu.cn. This work was supported by National Natural
Science Foundation of China (No. 11571203); Supported by the Programme of
Introducing Talents of Discipline to Universities of China (No.B12023). }
\and Xiaomin Shi\thanks{Corresponding author. Institute for Financial Studies,
Shandong University, Jinan 250100, China. Email: shixm@mail.sdu.edu.cn,
shixiaominhi@163.com. This work was supported by National Natural Science
Foundation of China (No. 11401091).}}
\date{}
\maketitle

\textbf{Abstract}. This paper concerns the recursive utility maximization
problem. We assume that the coefficients of the wealth equation and the
recursive utility are concave. Then some interesting and important cases with
nonlinear and nonsmooth coefficients satisfy our assumption. After given an
equivalent backward formulation of our problem, we employ the Fenchel-Legendre
transform and derive the corresponding variational formulation. By the convex
duality method, the primal \textquotedblleft sup-inf" problem is translated to
a dual minimization problem and the saddle point of our problem is derived.
Finally, we obtain the optimal terminal wealth. To illustrate our results,
three cases for investors with ambiguity aversion are explicitly worked out
under some special assumptions.

{\textbf{Key words}. } recursive utility optimization, convex duality,
nonsmooth coefficient, nonlinear equation, saddle point

\textbf{Mathematics Subject Classification (2010)} 60H10 93E20 49N15

\addcontentsline{toc}{section}{\hspace*{1.8em}Abstract}

\section{Introduction}

The problem of an agent who invests in a financial market so as to maximize
the utility of his terminal wealth is a topic that is being widely studied.
There have been many literatures on this issue with different viewpoints.

When the wealth equation is linear, the problem of maximizing the expected
utility of terminal wealth is well understood in a complete or constrained
financial market (refer to \cite{KS,KLS,CK}). Since then many works are
focused on this optimization problem with improved utility functions. For
example, Bian \cite{BMZ} and Westray, Zheng \cite{WZ1}, \cite{WZ2} studied
this kind of problem with nonsmooth utility functions. In order to deal with
model uncertainty, the so-called robust utility maximization problems are also
investigated widely. Quenez \cite{Qu} studied this problem in an incomplete
multiple-priors model. Schied \cite{Sc} explored the robust utility
maximization problem in a complete market under the existence of a
\textquotedblleft least favorable measure". Jin and Zhou \cite{JZ} studied
expected utility maximization problem as well as mean-variance problem when
the appreciation rates are only known to be in a certain convex closed set.
Other developments include the recursive utility maximization problems which
were investigated by Faidi et al \cite{FMM}, Matoussi, Xing \cite{MX} and
Epstein, Ji \cite{EJ1,EJ}.

But up to our knowledge, there are not many results related to this kind of
optimization problem with nonlinear wealth equations. Cvitanic and Cuoco
\cite{CC} explored the optimal consumption problem for a large investor whose
portfolio strategies can affect the instantaneous expected returns of the
assets. They show the existence of optimal policies by convex duality method
developed in \cite{CK,KLS}. Ji and Peng \cite{ji-peng} studied the continuous
time mean-variance problem with nonlinear wealth equation.\ El Karoui et al
\cite{EPQ} obtained a dynamic maximum principle for the optimization of
recursive utilities and characterized the optimal consumptions and portfolio
strategies via a forward-backward SDE system. Their method depends heavily on
the smoothness of the coefficients of the forward-backward SDE system.

As for the wealth equations, there are some interesting cases in which the
coefficients are nonlinear and nonsmooth. As shown in \cite{CC}, the wealth
equation of a large investor may be nonsmooth. The other well-known case is
that an investor is allowed to borrow money with a higher interest rate. As
for the recursive utilities, some important generators are also nonsmooth, for
instance, the K-ignorance case which was proposed by Chen and Epstein
\cite{CE}. The coefficients of the wealth equations and the recursive
utilities of the above cases are all concave. This motivate us to study the
recursive utility maximization problem with concave coefficients of both the
wealth equations and the recursive utilities in this paper.

We first give an equivalent backward formulation of our problem. This
\textquotedblleft backward formulation" was introduced by El Karoui, Peng and
Quenez \cite{EPQ} in order to solve a recursive utility optimization problem,
and employed by Ji and Peng \cite{ji-peng} to obtain a necessary condition for
a mean--variance portfolio selection problem with non-convex wealth equations.
For its application in stochastic control with state constraints, we refer the
reader to Ji, Zhou \cite{ji-zhou, ji-zhou-1}. El Karoui, Peng and Quenez
\cite{EPQ} took the terminal wealth as the control variable, and then used a
variational technique to obtain a stochastic maximum principle, i.e., a
first-order necessary condition that characterizes the terminal wealth. In our
context, we still take the terminal wealth as the control variable. But the
stochastic maximum principle approach does not work due to the nonsmoothness
of the coefficients. In order to overcome this difficulty, we assume that the
coefficients are concave and derive a variational formulation by the
Fenchel-Legendre transform of the coefficients which leads to a stochastic
game problem. Inspired by the convex duality method developed in Cvitanic and
Karatzas \cite{CK2}, we turn the primal \textquotedblleft sup-inf" problem to
a dual minimization problem and the saddle point of this game is derived. Then
we obtain the optimal terminal wealth and the optimal portfolio process can be
derived by the martingale representation theorem.

Three cases for investors with ambiguity aversion are provided to show the
applications of our results. In these cases, we specialize the generator of
the recursive utility as the K-ignorance case and the utility function of the
terminal wealth as $u(x)=\frac{1}{\alpha}x^{\alpha},\ 0<\alpha<1$. By the main
results in section 4, we characterize the saddle point via a quadratic BSDE
and obtain the optimal terminal wealth explicitly. Especially, for the large
investor case, we work out the explicit saddle point, the optimal wealth
process, the optimal portfolio strategies as well as the utility intensity process.

This paper is organized as follows. In section 2, we give the classical,
backward and variational formulation of the recursive utility maximization
problem. Our main results are obtained in section 3. In section 4, we study
three cases in which the investors are assumed to be ambiguity aversion
(K-ignorance). The saddle point and the optimal terminal wealth for each case
are derived explicitly.

\section{Formulation of the problem}

In this paper, we study the recursive utility maximization problem with
bankruptcy prohibition.

\subsection{The wealth process}

Let $W=(W^{1},...,W^{d})^{\prime}$ be a standard d-dimensional Brownian motion
defined on a filtered complete probability space $(\Omega,\mathcal{F}%
,\{\mathcal{F}_{t}\}_{t\geq0},P)$, where $\{\mathcal{F}_{t}\}_{t\geq0}$
denotes the natural filtration associated with the $d$-dimensional Brownian
motion $W$ and augmented.

Consider a financial market consisting of a riskless asset (the money market
instrument or bond) whose price is $S^{0}$ and d risky securities (the stocks)
whose prices are $S^{1},...,S^{d}$. An investor decides at time $t\in
\lbrack0,T]$ what amount $\pi_{t}^{i}$ of his wealth $X_{t}$ to invest in the
$i$th stock, $i=1,...,d$. The portfolio $\pi_{t}=(\pi_{t}^{1},...,\pi_{t}%
^{d})^{\prime}$ and $\pi_{t}^{0}:=X_{t}-\sum_{i=1}^{d}\pi_{t}^{i}$ are
$\mathcal{F}_{t}$-adapted. We suppose that the wealth process $X_{t}\equiv
X_{t}^{x,\pi}$ of the investor who is endowed with initial wealth $x>0$ is
governed by the following stochastic differential equation,
\begin{equation}%
\begin{cases}
dX_{t}=b(t,X_{t},\sigma_{t}^{\prime}\pi_{t})dt+\pi_{t}^{\prime}\sigma
_{t}dW_{t};\\
X_{0}=x,
\end{cases}
\label{wealth1}%
\end{equation}
where $b$ is a given function and the predictable and invertible process
$\sigma_{t}=\{\sigma_{t}^{ij}\}_{1\leq i,j\leq d}$ is the stock volatility.
$\sigma_{t}$ is assumed to be bounded, uniformly in $(t,\omega)\in
\lbrack0,T]\times\Omega$, and $\exists\varepsilon>0,\ \rho^{\prime}\sigma
_{t}\sigma_{t}^{\prime}\rho\geq\varepsilon||\rho||^{2},\ \forall\rho
\in\mathbb{R}^{d},\ t\in\lbrack0,T],$ $a.s.$

\begin{example}
\label{exam-1} The standard linear case.

The price processes $S_{t}^{0}$ and $S_{t}^{1},...,S_{t}^{d}$ are governed by
\begin{align}
&  dS_{t}^{0}=S_{t}^{0}r_{t}dt,\ S_{0}^{0}=s_{0};\label{stock-linear}\\
&  dS_{t}^{i}=S_{t}^{i}\big[b_{t}^{i}dt+\sum\limits_{i=1}^{d}\sigma_{t}%
^{ij}dW_{t}^{j}\big],\ S_{0}^{i}=s_{i}>0;\ i=1,...,d.\nonumber
\end{align}
All processes $r_{t},\ b_{t}=(b_{t}^{1},...,b_{t}^{d})^{\prime},\ \sigma
_{t}=\{\sigma_{t}^{ij}\}_{1\leq i,j\leq d},\ \sigma_{t}^{-1}$ are assumed to
be predictable and bounded, uniformly in $(t,\omega)\in\lbrack0,T]\times
\Omega.$ Then the wealth process $X_{t}$ satisfies the following linear
stochastic differential equation,
\[%
\begin{cases}
dX_{t}=\big(r_{t}X_{t}+\pi_{t}^{\prime}(b_{t}-r_{t}\mathbf{1})\big)dt+\pi
_{t}^{\prime}\sigma_{t}dW_{t},\\
X_{0}=x,
\end{cases}
\]
where $\mathbf{1}$ is the d-dimensional vector whose every component is 1. In
this example,
\begin{equation}
b(t,X_{t},\sigma_{t}^{\prime}\pi_{t})=r_{t}X_{t}+\pi_{t}^{\prime}(b_{t}%
-r_{t}\mathbf{1}). \label{driftsta}%
\end{equation}

\end{example}

\begin{example}
\label{exhigher} The borrowing rate $R_{t}$ is higher than the risk-free rate
$r_{t}$.

The stock prices are (\ref{stock-linear}). Now the borrowing rate $R_{t}$ is
higher than the risk-free rate $r_{t}$, i.e., $R_{t}\geq r_{t},$ $t\in
\lbrack0,T],$ $a.s.$ In this case, the wealth process becomes
\[%
\begin{cases}
dX_{t}=\big(r_{t}X_{t}+\pi_{t}^{\prime}(b_{t}-r_{t}\mathbf{1})-(R_{t}%
-r_{t})(X_{t}-\pi_{t}^{\prime}\mathbf{1})^{-}\big)dt+\pi_{t}^{\prime}%
\sigma_{t}dW_{t},\\
X_{0}=x.
\end{cases}
\]
In this example,
\begin{equation}
b(t,X_{t},\sigma_{t}^{\prime}\pi_{t})=r_{t}X_{t}+\pi_{t}^{\prime}(b_{t}%
-r_{t}\mathbf{1})-(R_{t}-r_{t})(X_{t}-\pi_{t}^{\prime}\mathbf{1})^{-}.
\label{drifthigh}%
\end{equation}

\end{example}

\begin{example}
\label{exam-2} A large investor case.

Cuoco and Cvitanic \cite{CC} considered the optimal portfolio choice problem
for a large investor whose portfolio strategies can affect the price processes
of the securities. In \cite{CC}, the price processes are given by
\begin{align*}
&  dS_{t}^{0}=S_{t}^{0}\big[r_{t}+l_{0}(X_{t},\pi_{t})\big]dt,\ S_{0}%
^{0}=s_{0};\\
&  dS_{t}^{i}=S_{t}^{i}\Big[\big(b_{t}^{i}+l_{i}(X_{t},\pi_{t})\big)dt+\sum
\limits_{j=1}^{d}\sigma_{t}^{ij}dW_{t}^{j}\Big],\ S_{0}^{i}=s_{i}%
>0,\ i=1,...,d,
\end{align*}
where $l_{i}:\mathbb{R}^{+}\times\mathbb{R}\rightarrow\mathbb{R}%
,\ i=0,1,...,d$ are some given functions which describe the effect of the
wealth and the strategies possessed by the large investor. The wealth process
is governed by
\[%
\begin{cases}
dX_{t}=\big(r_{t}X_{t}+(X_{t}-\pi_{t}^{\prime}\mathbf{1})l_{0}(X_{t},\pi
_{t})+\pi_{t}^{\prime}[b_{t}-r_{t}\mathbf{1}+l(X_{t},\pi_{t})]\big)dt+\pi
_{t}^{\prime}\sigma_{t}dW_{t},\\
X_{0}=x.
\end{cases}
\]
In this example,
\[
b(t,X_{t},\sigma_{t}^{\prime}\pi_{t})=r_{t}X_{t}+(X_{t}-\pi_{t}^{\prime
}\mathbf{1})l_{0}(X_{t},\pi_{t})+\pi_{t}^{\prime}[b_{t}-r_{t}\mathbf{1}%
+l(X_{t},\pi_{t})].
\]
Cuoco and Cvitanic \cite{CC} also gave the following more specific example.
For $x\in\mathbb{R}$,
\begin{equation}
sgn(x):=%
\begin{cases}
\frac{|x|}{x},\ \ \text{if}\ \ x\neq0;\\
0,\ \ \ \ \text{otherwise}.
\end{cases}
\end{equation}
Set $\varepsilon=(\varepsilon_{1},...,\varepsilon_{d})^{\prime}$, $sgn(\pi
_{t})=(sgn(\pi_{t}^{1}),...,sgn(\pi_{t}^{d}))^{\prime}$ and $\varepsilon
\otimes sgn(\pi_{t})=(\varepsilon_{1}sgn(\pi_{t}^{1}),...,\varepsilon
_{d}sgn(\pi_{t}^{d}))^{\prime}$ where $\varepsilon_{i}$ are given small
positive numbers.

Consider $l_{0}(X_{t},\pi_{t})=0$ and $\ l_{i}(X_{t},\pi_{t})=-\varepsilon
_{i}sgn(\pi_{t}^{i}),\ i=1,...,d$. Then we have
\begin{equation}
b(t,X_{t},\sigma_{t}^{\prime}\pi_{t})=r_{t}X_{t}+\pi_{t}^{\prime}(b_{t}%
-r_{t}\mathbf{1}-\varepsilon\otimes sgn(\pi_{t})). \label{driftlar}%
\end{equation}
For this specific large investor model, longing the ith risky security
depresses its expected return while shorting it increases its expected return
as explained in Cuoco and Cvitanic \cite{CC}.
\end{example}

We introduce the following spaces:
\[%
\begin{array}
[c]{l}%
L^{2}(\Omega,\mathcal{F}_{T},P;R)=\Big\{\xi:\Omega\rightarrow R\big|\xi
\mbox { is }\mathcal{F}_{T}\mbox{-measurable, and }E|\xi|^{2}<\infty\Big\},\\
M^{2}(0,T;R^{d})=\Big\{\phi:[0,T]\times\Omega\rightarrow R^{d}\big|(\phi
_{t})_{0\leq t\leq T}\mbox{ is }\{\mathcal{F}_{t}\}_{t\geq0}%
\mbox{-progressively measurable process,}\\
\mbox{  \ \ \ \  and }||\phi||^{2}=E\int_{0}^{T}|\phi_{t}|^{2}dt<\infty
\Big\},\\
S^{2}(0,T;R^{d})=\Big\{\phi:[0,T]\times\Omega\rightarrow R\big|(\phi
_{t})_{0\leq t\leq T}\mbox{ is }\{\mathcal{F}_{t}\}_{t\geq0}%
\mbox{-progressively measurable process,}\\
\mbox{  \ \ \ \  and }||\phi||_{S}^{2}=E[\sup\limits_{0\leq t\leq T}|\phi
_{t}|^{2}]<\infty\Big\}.
\end{array}
\]

For notational simplicity, we will often write $L^{2}$, $M^{2}$ and $S^{2}$
instead of $L^{2}(\Omega, \mathcal{F}_{T}, P; R)$, $M^{2}(0,T;R^{d})$ and
$S^{2}(0,T;R)$ respectively.

Let $b(\omega,t,X,\pi):\Omega\times\lbrack0,T]\times\mathbb{R}\times{R}%
^{d}\rightarrow\mathbb{R}$.

\begin{assumption}
\label{drift} $b(\omega,t,X,\pi):\Omega\times\lbrack0,T]\times\mathbb{R}%
\times\mathbb{R}^{d}\rightarrow\mathbb{R}$ is $\{\mathcal{F}_{t}\}_{t\geq0}%
$-progressively measurable for any $(X,\pi)\in\mathbb{R}\times\mathbb{R}^{d}$
and\newline(i) There exists a constant $C_{1}\geq0$ such that
\[
\big|b(\omega,t,X^{1},\pi^{1})-b(\omega,t,X^{2},\pi^{2})\big|\leq
C_{1}(\big|X^{1}-X^{2}\big|+\big|\pi^{1}-\pi^{2}\big|),\ \forall
(t,\omega,X^{1},X^{2},\pi^{1},\pi^{2})\in\Omega\times\lbrack0,T]\times
\mathbb{R}\times\mathbb{R}\times\mathbb{R}^{d}\times\mathbb{R}^{d}.
\]
(ii) $b(t,0,0)\geq0,\ t\in\lbrack0,T],\ a.s.$ and $E\int_{0}^{T}%
b^{2}(t,0,0)dt<+\infty.$\newline(iii) \text{The function} \ $(X,\pi)\mapsto
b(\omega,t,X,\pi)$ \text{is concave for all} \ $(\omega,t)\in\Omega
\times\lbrack0,T]$.
\end{assumption}

\subsection{The recursive utility}

In the time-additive expected utility maximization models, one can not
separate the risk aversion and intertemporal substitution. To overcome this
intertwine, Duffie and Epstein \cite{DE} introduced the stochastic recursive
utility in the continuous time. In El Karoui, Peng and Qeuenz \cite{EPQ1}, the
stochastic recursive utility can be formulated in a more general form by
backward stochastic differential equation (BSDE for short):
\begin{equation}
Y_{t}=u(X_{T})+\int_{t}^{T}f(s,Y_{s},Z_{s})ds-\int_{t}^{T}Z_{s}^{\prime}%
dW_{s}. \label{BSDE}%
\end{equation}

We need the following assumptions.

\begin{assumption}
\label{assf} $f:\Omega\times\lbrack0,T]\times\mathbb{R}\times\mathbb{R}%
^{d}\rightarrow\mathbb{R}$ is $\{\mathcal{F}_{t}\}_{t\geq0}$-progressively
measurable for any $(y,z)\in\mathbb{R}\times\mathbb{R}^{d}$, and
satisfies\newline(i) There exists a constant $C\geq0$ such that
\[
\big|f(\omega,t,Y_{1},Z_{1})-f(\omega,t,Y_{2},Z_{2})\big|\leq C(\big|Y_{1}%
-Y_{2}\big|+\big|Z_{1}-Z_{2}\big|),\ \forall(\omega,t,Y_{1},Y_{2},Z_{1}%
,Z_{2})\in\Omega\times\lbrack0,T]\times\mathbb{R}\times\mathbb{R}%
\times\mathbb{R}^{d}\times\mathbb{R}^{d};
\]
(ii) $f$ is continuous in $t$ and $E\int_{0}^{T}f^{2}(t,0,0)dt<+\infty
;$\newline(iii) \text{The function} \ $(Y,Z)\mapsto f(\omega,t,Y,Z)$ \text{is
concave for all} \ $(\omega,t)\in\Omega\times\lbrack0,T]$.
\end{assumption}

\begin{assumption}
\label{assu} $u:(0,\infty)\rightarrow\mathbb{R}$ is strictly increasing,
strictly concave and of class $C^{2}$, and satisfies\newline(i) Inada
condition:
\[
u^{\prime}(0+):=\lim_{x\rightarrow0+}u^{\prime}(x)=\infty,u^{\prime}%
(\infty):=\lim_{x\rightarrow\infty}u^{\prime}(x)=0;
\]
(ii) $\exists k_{1},k_{2}\geq0,$ $p\in(0,1)$ such that $|u(x)|\leq k_{1}%
+k_{2}\frac{x^{p}}{p}$;\newline(iii) $u(0):=\lim\limits_{x\rightarrow0^{+}%
}u(x)>-\infty$; $u(\infty):=\lim\limits_{x\rightarrow\infty}u(x)=\infty$.
\end{assumption}

\subsection{Classical formulation}

We consider that an investor chooses a portfolio strategy so as to%
\begin{align}
&  \text{ }\mathrm{Maximize}\ Y_{0}^{x,\pi},\label{optm}\\
&  \text{ }s.t.%
\begin{cases}
X_{t}\geq0,\\
\pi\in M^{2},\\
(X,\pi)\ \ \ \mathrm{satisfies\ \ \ }(\ref{wealth1}),\\
(Y,Z)\ \ \ \mathrm{satisfies\ \ \ }(\ref{BSDE}),
\end{cases}
\nonumber
\end{align}
where $X_{t}\geq0$ describes that no-bankruptcy is required.

\begin{definition}
A portfolio $\pi$ is said to be admissible if $\pi\in M^{2}$ and the
corresponding wealth process $X_{t}\geq0,\ t\in\lbrack0,T],a.s.$
\end{definition}

Given the initial wealth $x>0$, denote by $\mathcal{\bar{A}}(x)$ the set of an
investor's admissible portfolio strategies, that is
\begin{equation}
{\mathcal{\bar{A}}}(x)\equiv\mathcal{\bar{A}}(x;0,T)=\Big\{\pi\mid\pi\in
M^{2},X_{t}^{x,\pi}\geq0,\ t\in\lbrack0,T],\ a.s.\Big\}. \label{admispi}%
\end{equation}

Thus, (\ref{optm}) can be written as:%
\begin{align*}
&  \mathrm{Maximize}\ Y_{0}^{x,\pi},\\
&  s.t.%
\begin{cases}
\pi\in\mathcal{\bar{A}}(x),\\
(X,\pi)\ \ \ \mathrm{satisfies\ \ \ }(\ref{wealth1}),\\
(Y,Z)\ \ \ \mathrm{satisfies\ \ \ }(\ref{BSDE}),
\end{cases}
\end{align*}

\subsection{Backward formulation}

In this subsection, we give an equivalent backward formulation of the above
optimization problem (\ref{optm}). This backward formulation is founded in
\cite{EPQ,ji-peng, ji-zhou, ji-zhou-1}.

Set
\begin{align*}
q_{t}  &  =\sigma_{t}^{\prime}\pi_{t},\text{ }a.s.,\text{ }\forall t\in
\lbrack0,T],\\
U  &  =\{\xi\big|\xi\in L^{2}\ \text{and}\ \xi\geq0\}.
\end{align*}
Since $\sigma_{t}$ is invertible, $q_{t}$ can be regarded as the control
variable instead of $\pi_{t}$. Notice that selecting $q$\ is equivalent to
selecting the terminal wealth $X_{T}$ by the existence and uniqueness result
of BSDEs (refer to Theorem 2.1 in \cite{EPQ1}). Hence the wealth equation
(\ref{wealth1}) can be rewritten as
\begin{equation}
\left\{
\begin{array}
[c]{l}%
-dX_{t}=-b(t,X_{t},q_{t})dt-q_{t}^{\prime}dW_{t},\\
X_{T}=\xi,
\end{array}
\right.  \label{wealth1b}%
\end{equation}
where the terminal wealth $\xi$\ is the \textquotedblleft control" to be chosen from $U$. Note
that we will require that the solution $X$ of (\ref{wealth1b}) at time $0$
equals the initial wealth $x$.

If we take the terminal wealth as control variable, the recursive utility
process can be written as:
\begin{equation}%
\begin{cases}
-dY_{t}=f(t,Y_{t},Z_{t})dt-Z_{t}^{\prime}dW_{t},\\
Y_{T}=u(\xi).
\end{cases}
\label{wealth1bb}%
\end{equation}

Assumption \ref{assu} guarantees that $u(\xi)\in L^{2}$ for any $\xi\in U$. By
the existence and uniqueness result of BSDEs, we know that for any $\xi\in U$,
there exists a unique solution $(X_{t},q_{t})$ (resp. $(Y_{t},Z_{t})$) of
(\ref{wealth1b}) (resp. (\ref{wealth1bb})). By the comparison theorem of
BSDEs, Assumption \ref{drift} and the nonnegative terminal wealth keeps the
wealth process be nonnegative all the time. Usually, we denote the solution
$Y$ of (\ref{wealth1bb}) at tome $0$ by $Y_{0}^{\xi}$.

This gives rise to the following optimization problem:%
\begin{equation}%
\begin{array}
[c]{l}%
\text{ }\mathrm{Maximize}\ J(\xi):=Y_{0}^{\xi},\\
s.t.%
\begin{cases}
\xi\in U,\\
X_{0}=x,\\
(X,q)\text{ and }(Y,Z)\text{ satisfy }(\ref{wealth1b})\text{ and
}(\ref{wealth1bb})\text{ respectively}.
\end{cases}
\end{array}
\label{optmb}%
\end{equation}

\begin{definition}
A random variable $\xi\in U$ is called feasible for the initial wealth $x$ if
and only if $X_{0}=x$. We will denote by $\mathcal{A}(x)$ the set of all
feasible $\xi$ for the initial wealth $x$.
\end{definition}

It is clear that the original problem (\ref{optm}) is equivalent to the
auxiliary one (\ref{optmb}). Hence, hereafter we focus ourselves on solving
problem (\ref{optmb}). The advantage of doing this is that, since $\xi$ is the
control variable, the state constraint in (\ref{optm}) becomes a control
constraint in (\ref{optmb}), whereas it is well known in control theory that a
control constraint is much easier to deal with than a state constraint. The
cost of this approach is that the original initial condition $X_{0}=x$ becomes
a constraint.

\subsection{Variational formulation}

Let $\tilde{b}(\omega,t,\mu,\nu)$ be the Fenchel-Legendre transform of $b$:
\begin{equation}
\tilde{b}(\omega,t,\mu,\nu)=\sup_{(x,q)\in\mathbb{R}\times\mathbb{R}^{d}%
}\big[b(\omega,t,x,q)-x\mu-q^{\prime}\nu\big],\ (\mu,\nu)\in\mathbb{R}%
\times\mathbb{R}^{d}.
\end{equation}
The effective domain of $\tilde{b}$ is
\[
\mathcal{D}_{\tilde{b}}:=\{(\omega,t,\mu,\nu)\in\Omega\times\lbrack
0,T]\times\mathbb{R}\times{\mathbb{R}}^{d}\big|\ \tilde{b}(\omega,t,\mu
,\nu)<+\infty\}.
\]
As was shown in \cite{EPQ1}, the $(\omega,t)$-section of $\mathcal{D}%
_{\tilde{b}}$, denoted by $\mathcal{D}_{\tilde{b}}^{(\omega,t)}$ is included
in the bounded domain
\[
B^{\prime}:=[-C_{1},C_{1}]^{d+1}\subset\mathbb{R}\times\mathbb{R}^{d},
\]
where $C_{1}$ is the Lipschitz constant of $b$. The following duality relation
is due to the concavity of $b$,
\begin{equation}
b(\omega,t,x,q)=\inf_{(\mu,\nu)\in\mathcal{D^{\prime}}_{\tilde{b}}%
^{(\omega,t)}}\big[\tilde{b}(\omega,t,\mu,\nu)+x\mu+q^{\prime}\nu\big].
\end{equation}
Set
\[
\mathcal{B^{\prime}}=\big\{(\mu,\nu)\Big|(\mu,\nu)\ \text{is}\ \{\mathcal{F}%
_{t}\}_{t\geq0}\text{-progressively measurable and}\ B^{\prime}\text{-valued}%
\ \text{and}\ E\int_{0}^{T}\tilde{b}(\omega,t,\mu_{t},\nu_{t})^{2}%
dt<+\infty\big\}.
\]

Let $F(\omega,t,\beta,\gamma)$ be the Fenchel-Legendre transform of $f$:
\begin{equation}
F(\omega,t,\beta,\gamma)=\sup_{(y,z)\in\mathbb{R}\times\mathbb{R}^{d}%
}\big[f(\omega,t,y,z)-y\beta-z^{\prime}\gamma\big],\ (\beta,\gamma
)\in\mathbb{R}\times\mathbb{R}^{d}.
\end{equation}
The effective domain of $F$ is
\[
\mathcal{D}_{F}:=\{(\omega,t,\beta,\gamma)\in\Omega\times\lbrack
0,T]\times\mathbb{R}\times{\mathbb{R}}^{d}\big|F(\omega,t,\beta,\gamma
)<+\infty\}.
\]
Similarly, the $(\omega,t)$-section of $\mathcal{D}_{F}$, denoted by
$\mathcal{D}_{F}^{(\omega,t)}$ is included in the bounded domain
\[
B:=[-C,C]^{d+1}\subset\mathbb{R}\times\mathbb{R}^{d},
\]
where $C$ is the Lipschitz constant of $f$. We have the duality relation by
the concavity of $f$,
\begin{equation}
f(\omega,t,y,z)=\inf_{(\beta,\gamma)\in\mathcal{D}_{F}^{(\omega,t)}%
}\big[F(\omega,t,\beta,\gamma)+y\beta+z^{\prime}\gamma\big].
\end{equation}
Set
\[
\mathcal{B}=\big\{(\beta,\gamma)\Big |(\beta,\gamma)\ \text{is}\ \{\mathcal{F}%
_{t}\}_{t\geq0}\text{-progressively measurable and B-valued}\ \text{and}%
\ E\int_{0}^{T}F(t,\beta_{t},\gamma_{t})^{2}dt<+\infty\big\}.
\]

\begin{assumption}
\label{bound} The functions $\tilde b$ and $F$ are bounded on their effective domains.
\end{assumption}

\begin{lemma}
Under Assumption \ref{drift}, \ref{assf} and \ref{bound}, $\mathcal{B^{\prime
}}$ and $\mathcal{B}$ are convex sets and are closed under almost sure convergence.
\end{lemma}

\noindent\textbf{Proof:} The convexity of $\mathcal{B^{\prime}}$ and
$\mathcal{B}$ comes from the convexity of $\tilde{b}$ and $F$. Assumption
\ref{bound} with Fatou's Lemma guarantees the closeness. $\ \ \ \ \ \Box$
%\begin{assumption}
%$\mathcal{B'}\bigcap\mathcal{B}=\emptyset.$
%\end{assumption}

For any $(\mu,\nu)\in\mathcal{B}^{\prime}$, define
\[
b^{\mu,\nu}(t,y,z)=\tilde{b}(t,\mu_{t},\nu_{t})+x\mu_{t}+\pi^{\prime}\nu_{t},
\]
and denote by $(X^{\mu,\nu},q^{\mu,\nu})$ the unique solution to the linear
BSDE (\ref{wealth1b}) associated to $b^{\mu,\nu}$. For any $(\beta,\gamma
)\in\mathcal{B}$, define
\[
f^{\beta,\gamma}(t,y,z)=F(t,\beta_{t},\gamma_{t})+y\beta_{t}+z^{\prime}%
\gamma_{t},
\]
and denote by $(Y^{\beta,\gamma},Z^{\beta,\gamma})$ the unique solution to the
linear BSDE (\ref{wealth1bb}) associated to $f^{\beta,\gamma}$.

For $0\leq t\leq s\leq T$, set%
\begin{align*}
N_{t,s}^{\mu,\nu}  &  =e^{-\int_{t}^{s}(\mu_{r}+\frac{1}{2}\parallel\nu
_{r}\parallel^{2})dr-\int_{t}^{s}\nu_{r}^{\prime}dW_{r}},\\
\Gamma_{t,s}^{\beta,\gamma}  &  =e^{\int_{t}^{s}(\beta_{r}-\frac{1}%
{2}\parallel\gamma_{r}\parallel^{2})dr+\int_{t}^{s}\gamma_{r}^{\prime}dW_{r}}.
\end{align*}
By the method similar to Proposition 3.4 in \cite{EPQ1}, we have the following
variational formulation of $X_{t}$ and $Y_{t}$.

\label{rep} Under Assumption \ref{drift} and \ref{assf}, the solutions
$(X_{t},q_{t})$ and $(Y_{t},Z_{t})$ of (\ref{wealth1b}) and (\ref{wealth1bb})
can be represented as
\begin{align*}
&  X_{t}=\underset{(\mu,\nu)\in\mathcal{B^{\prime}}}{ess\sup}X_{t}^{\mu,\nu
},\text{ }a.s.,\\
&  Y_{t}=\underset{(\beta,\gamma)\in\mathcal{B}}{ess\inf}Y_{t}^{\beta,\gamma
},\ a.s.,
\end{align*}
where
\begin{align*}
&  X_{t}^{\mu,\nu}=E\big[-\int_{t}^{T}N_{t,s}^{\mu,\nu}\tilde{b}(s,\mu
,\nu)ds+N_{t,T}^{\mu,\nu}\xi|\mathcal{F}_{t}\big],\\
&  Y_{t}^{\beta,\gamma}=E\big[\int_{t}^{T}\Gamma_{t,s}^{\beta,\gamma}%
F(s,\beta_{s},\gamma_{s})ds+\Gamma_{t,T}^{\beta,\gamma}u(\xi)|\mathcal{F}%
_{t}\big],
\end{align*}

Especially, we have
\begin{align*}
&  X_{0}=\sup\limits_{(\mu,\nu)\in\mathcal{B^{\prime}}}E\big[-\int_{0}%
^{T}N_{0,s}^{\mu,\nu}\tilde{b}(s,\mu_{s},\nu_{s})ds+N_{0,T}^{\mu,\nu}%
\xi\big],\\
&  Y_{0}=\inf\limits_{(\beta,\gamma)\in\mathcal{B}}E\big[\int_{0}^{T}%
\Gamma_{0,s}^{\beta,\gamma}F(s,\beta_{s},\gamma_{s})ds+\Gamma_{0,T}%
^{\beta,\gamma}u(\xi)\big].
\end{align*}
By the above analysis,
\begin{equation}
\mathcal{A}(x)=\{\xi\in U\big|\sup\limits_{(\mu,\nu)\in\mathcal{B^{\prime}}%
}E\big[-\int_{0}^{T}N_{0,s}^{\mu,\nu}\tilde{b}(s,\mu_{s},\nu_{s}%
)ds+N_{0,T}^{\mu,\nu}\xi\big]=x\}. \label{constraint-xi}%
\end{equation}

Now our problem (\ref{optmb}) is equivalent to the following problem:%
\begin{align}
\mathrm{Maximize}\ J(\xi)  &  =Y_{0}^{\xi}=\inf\limits_{(\beta,\gamma
)\in\mathcal{B}}E\big[\int_{0}^{T}\Gamma_{0,s}^{\beta,\gamma}F(s,\beta
_{s},\gamma_{s})ds+\Gamma_{0,T}^{\beta,\gamma}u(\xi
)\big],\label{varformulation}\\
s.t.\ \ \xi &  \in\mathcal{A}(x).\nonumber
\end{align}
\

It is essentially a robust optimization problem. Define the \textquotedblleft
max-min" quantity
\begin{equation}
\underline{V}(x)=\sup_{\xi\in\mathcal{A}(x)}\inf\limits_{(\beta,\gamma
)\in\mathcal{B}}E\big[\int_{0}^{T}\Gamma_{0,s}^{\beta,\gamma}F(s,\beta
_{s},\gamma_{s})ds+\Gamma_{0,T}^{\beta,\gamma}u(\xi)\big].
\end{equation}
It is dominated by its \textquotedblleft min-max" counterpart
\[
\bar{V}(x)=\inf\limits_{(\beta,\gamma)\in\mathcal{B}}\sup_{\xi\in
\mathcal{A}(x)}E\big[\int_{0}^{T}\Gamma_{0,s}^{\beta,\gamma}F(s,\beta
_{s},\gamma_{s})ds+\Gamma_{0,T}^{\beta,\gamma}u(\xi)\big].
\]
If we can find $(\hat{\beta},\hat{\gamma},\hat{\xi})\in\mathcal{B}%
\times\mathcal{A}(x)$ such that
\begin{equation}
\underline{V}(x)=E\big[\int_{0}^{T}\Gamma_{0,s}^{\hat{\beta},\hat{\gamma}%
}F(s,\hat{\beta}_{s},\hat{\gamma}_{s})ds+\Gamma_{0,T}^{\hat{\beta},\hat
{\gamma}}u(\hat{\xi})\big]=\bar{V}(x), \label{game1}%
\end{equation}
then the optimal solution of problem (\ref{varformulation}) is $\hat{\xi}$.

\section{Main results}

Denote the inverse of the marginal utility function $u^{\prime}(\cdot)$ by
$I(\cdot)$. The convex dual
\begin{equation}
\tilde{u}(\zeta):=\max_{x>0}[u(x)-\zeta x]=u(I(\zeta))-\zeta I(\zeta
),\ \zeta>0.
\end{equation}

For $0<\zeta<\infty$, introduce the value functions
\begin{equation}
\tilde{V}(\zeta)=\inf_{\substack{(\beta,\gamma)\in\mathcal{B}\\(\mu,\nu
)\in\mathcal{B^{\prime}}}}E\big[\int_{0}^{T}(\Gamma_{0,s}^{\beta,\gamma
}F(s,\beta_{s},\gamma_{s})+\zeta N_{0,s}^{\mu,\nu}\tilde{b}(s,\mu_{s},\nu
_{s}))ds+\Gamma_{0,T}^{\beta,\gamma}\tilde{u}\big(\zeta\frac{N_{0,T}^{\mu,\nu
}}{\Gamma_{0,T}^{\beta,\gamma}}\big)\big] \label{dua2}%
\end{equation}
and
\begin{align}
V_{\ast}(x)  &  =\inf_{\substack{(\beta,\gamma)\in\mathcal{B}\\(\mu,\nu
)\in\mathcal{B^{\prime}}\\\zeta>0}}E\big[\int_{0}^{T}(\Gamma_{0,s}%
^{\beta,\gamma}F(s,\beta_{s},\gamma_{s})+\zeta N_{0,s}^{\mu,\nu}\tilde
{b}(s,\mu_{s},\nu_{s}))ds+\Gamma_{0,T}^{\beta,\gamma}\tilde{u}\big(\zeta
\frac{N_{0,T}^{\mu,\nu}}{\Gamma_{0,T}^{\beta,\gamma}}\big)+\zeta
x\big]\nonumber\label{dua11}\\
&  =\inf_{\zeta>0}[\tilde{V}(\zeta)+\zeta x].
\end{align}

\begin{lemma}
\label{hatnu11} Under Assumption \ref{assf}, \ref{assu} and \ref{bound}, for
any given $\zeta>0$, there exist pairs $(\hat{\beta},\hat{\gamma})=(\hat
{\beta}_{\zeta},\hat{\gamma}_{\zeta})\in\mathcal{B}$ and $(\hat{\mu},\hat{\nu
})=(\hat{\mu}_{\zeta},\hat{\nu}_{\zeta})\in\mathcal{B^{\prime}}$ which attain
the infimum in (\ref{dua2}).
\end{lemma}

\noindent\textbf{Proof:} By the boundedness of $\mathcal{B}$ and
$\mathcal{B^{\prime}}$, the sets $\mathcal{\tilde{B^{\prime}}}=\{N_{0,T}%
^{\mu,\nu}:(\mu,\nu)\in\mathcal{B^{\prime}}\}$ and $\mathcal{\tilde{B}%
}=\{\Gamma_{0,T}^{\beta,\gamma}:(\beta,\gamma)\in\mathcal{B}\}$ are bounded in
$L^{p}$ for any $p\geq1$. So $(\mu,\nu)\in\mathcal{B^{\prime}}$ (resp.
$(\beta,\gamma)\in\mathcal{B}$) is uniquely determined by $N_{0,T}^{\mu,\nu
}\in\mathcal{\tilde{B}^{\prime}}$ (resp. $\Gamma_{0,T}^{\beta,\gamma}%
\in\mathcal{\tilde{B}}$) (up to a.e. a.s. equivalence). Then,%
\[
\tilde{V}(\zeta)=\inf_{\substack{\Gamma_{0,T}^{\beta,\gamma}\in\mathcal{\tilde
{B}}\\N_{0,T}^{\mu,\nu}\in\mathcal{\tilde{B}^{\prime}}}}E\big[\int_{0}%
^{T}(\Gamma_{0,s}^{\beta,\gamma}F(s,\beta_{s},\gamma_{s})+\zeta N_{0,s}%
^{\mu,\nu}\tilde{b}(s,\mu_{s},\nu_{s}))ds+\Gamma_{0,T}^{\beta,\gamma}\tilde
{u}\big(\zeta\frac{N_{0,T}^{\mu,\nu}}{\Gamma_{0,T}^{\beta,\gamma}%
}\big)\big],\ 0<\zeta<\infty.
\]

Note that the function
\[
(x,y)\mapsto x\tilde{u}(\zeta\frac{y}{x}),\ \zeta>0
\]
is convex (not strictly). Then, by the method similar to that in \cite{CC}
(Theorem 3), we obtain the existence of $(\hat{\beta},\hat{\gamma})$ and
$(\hat{\mu},\hat{\nu})$. $\ \ \ \ \ \Box$

\begin{lemma}
\label{hatzeta1} Under Assumption \ref{drift}, \ref{assf} and \ref{assu},
%\[
%E\big[\int_0^T(\Gamma_{0,s}^{\beta,\gamma}F(s,\beta_s,\gamma_s)+\zeta  N_{\mu, \nu}(s)\tilde b(s,\mu_s, \nu_s)) ds+\Gamma_{0,T}^{\beta,\gamma}\tilde u\big(\zeta\frac{ N_{\mu, \nu}(T)}{\Gamma_{0,T}^{\beta,\gamma}}\big)\big]<\infty,\  \forall \zeta>0,\  \forall(\beta,\gamma)\in\mathcal{B}, \ \forall (\mu,\nu)\in\mathcal{B'}.
%\]
for any given $x>0$, there exists a number $\hat{\zeta}_{x}\in(0,\infty)$
which attains the infimum of $V_{\ast}(x)=\inf\limits_{\zeta>0}[\tilde
{V}(\zeta)+\zeta x]$.
\end{lemma}

\noindent\textbf{Proof:} By Assumption \ref{drift}, we know that for any
$s\in\lbrack0,T]$ and $(\mu,\nu)\in\mathcal{B}^{\prime}$, $\tilde{b}(s,\mu
_{s},\nu_{s})\geq0,\ a.s.$ Then, $\forall(\beta,\gamma)\in\mathcal{B}$ and
$(\mu,\nu)\in\mathcal{B}^{\prime},$
\begin{align*}
&\ \ \ \   E\big[\int_{0}^{T}(\Gamma_{0,s}^{\beta,\gamma}F(s,\beta_{s},\gamma
_{s})+\zeta N_{0,s}^{\mu,\nu}\tilde{b}(s,\mu_{s},\nu_{s}))ds+\Gamma
_{0,T}^{\beta,\gamma}\tilde{u}\big(\zeta\frac{N_{0,T}^{\mu,\nu}}{\Gamma
_{0,T}^{\beta,\gamma}}\big)\big]\\
&  \geq E\big[\int_{0}^{T}\Gamma_{0,s}^{\beta,\gamma}F(s,\beta_{s},\gamma
_{s})ds+\Gamma_{0,T}^{\beta,\gamma}\tilde{u}\big(\zeta\frac{N_{0,T}^{\mu,\nu}%
}{\Gamma_{0,T}^{\beta,\gamma}}\big)\big].
\end{align*}
By Assumption \ref{bound}, there exists a constant $M_{1}$ such that
$\forall(\beta,\gamma)\in\mathcal{B}$,
\[
E\int_{0}^{T}\Gamma_{0,s}^{\beta,\gamma}F(s,\beta_{s},\gamma_{s})ds\geq
M_{1}.
\]
Due to the monotonicity of $\tilde{u}$ and the convexity of $(x,y)\mapsto
x\tilde{u}(\zeta\frac{y}{x})$, we deduce the following inequality by Jensen's
inequality%
\[
E\big[\Gamma_{0,T}^{\beta,\gamma}\tilde{u}\big(\zeta\frac{N_{0,T}^{\mu,\nu}%
}{\Gamma_{0,T}^{\beta,\gamma}}\big)\big]\geq M_{2}\tilde{u}(M_{3}\zeta),\text{
}\forall(\beta,\gamma)\in\mathcal{B}\text{ and }(\mu,\nu)\in\mathcal{B}%
^{\prime},
\]
where the constants $M_{2}>0$, $M_{3}>0$ depend on the bound of $F$ and the
Lipschitz constants of $b$, $f$.

Hence $\tilde{V}(\zeta)\geq M_{1}+M_{2}\tilde{u}(M_{3}\zeta),\ \forall\zeta
>0$. By Assumption \ref{assu},
\begin{equation}
\tilde{V}(0):=\lim\limits_{\zeta\rightarrow0^{+}}\tilde{V}(\zeta)\geq
M_{1}+M_{2} \lim\limits_{\zeta\rightarrow0^{+}}\tilde{u}(M_{3}\zeta
)=M_{1}+M_{2}u(\infty)=\infty.
\end{equation}

From Lemma 4.2 in \cite{KLSX} and Assumption \ref{assu}, we know
\[
\tilde{u}(\infty):=\lim\limits_{\zeta\rightarrow\infty}\tilde{u}%
(\zeta)=u(0)>-\infty.
\]
Then we have
\begin{equation}
\lim\limits_{\zeta\rightarrow\infty}[\tilde{V}(\zeta)+\zeta x]\geq
\lim\limits_{\zeta\rightarrow\infty}[M_{1}+M_{2}\tilde{u}(M_{3}\zeta)+\zeta
x]=\infty,\ \forall x>0.
\end{equation}

So there exists a number $\hat{\zeta}_{x}\in(0,\infty)$ which attains the
infimum of $V_{\ast}(x)$. $\ \ \ \ \ \ \Box$

%Step 2: Notice that $\frac{N_{\hat\mu, \hat\nu}(T)}{\Gamma_{0,T}^{\hat\beta, \hat\gamma}}<+\infty, a.s. \ \forall (\beta, \gamma)\in\mathcal{B}$, because we have assumed that $\mu(\cdot)$ is bounded. Then
%\begin{align*}
%\tilde V'(+\infty):=\lim_{\zeta\rightarrow+\infty}\tilde V'(\zeta)&=-\lim_{\zeta\rightarrow\infty}E\big[N_{\hat\mu, \hat\nu}(T) I\big(\zeta\frac{N_{\hat\mu, \hat\nu}(T)}{\Gamma_{0,T}^{\hat\beta, \hat\gamma}}\big)\big]+E\int_0^T N_{\hat\mu, \hat\nu}(s)\tilde b(s,\hat\mu_s, \hat\nu_s)ds\\
%&=E\int_0^T N_{\hat\mu, \hat\nu}(s)\tilde b(s,\hat\mu_s, \hat\nu_s)ds\geq 0,
%\end{align*}
%and
%\[
%\tilde V'(0):=\lim_{\zeta\rightarrow 0+}\tilde V'(\zeta)=-\lim_{\zeta\rightarrow 0+}E\big[N_{\hat\mu, \hat\nu}(T) I\big(\zeta\frac{N_{\hat\mu, \hat\nu}(T)}{\Gamma_{0,T}^{\hat\beta, \hat\gamma}}\big)\big]+E\int_0^T N_{\hat\mu, \hat\nu}(s)\tilde b(s,\hat\mu_s, \hat\nu_s)ds=-\infty.
%\]
%And the infimum of $V_*(x)=\inf\limits_{\zeta>0}[\tilde V(\zeta)+\zeta x]$ must satisfy $\tilde V'(\zeta)=-x\in(-\infty, 0$. This complete the proof.   $\ \ \ \ \ \Box$

\begin{lemma}
\label{hatnuzeta1} Under Assumption \ref{drift}, \ref{assf}, \ref{assu} and
\ref{bound},
\[
V_{\ast}(x)=E\big[\int_{0}^{T}(\Gamma_{0,s}^{\hat{\beta},\hat{\gamma}}%
F(s,\hat{\beta}_{s},\hat{\gamma}_{s})+\hat{\zeta}N_{0,s}^{\hat{\mu},\hat{\nu}%
}\tilde{b}(s,\hat{\mu}_{s},\hat{\nu}_{s}))ds+\Gamma_{0,T}^{\hat{\beta}%
,\hat{\gamma}}\tilde{u}\big(\hat{\zeta}\frac{N_{0,T}^{\hat{\mu},\hat{\nu}}%
}{\Gamma_{0,T}^{\hat{\beta},\hat{\gamma}}}\big)+\hat{\zeta}x\big]
\]
with $\hat{\zeta}=\hat{\zeta}_{x}$ as in lemma \ref{hatzeta1} and $(\hat
{\beta},\hat{\gamma})=(\hat{\beta}_{\hat{\zeta}},\hat{\gamma}_{\hat{\zeta}%
})\in\mathcal{B}$, \ $(\hat{\mu},\hat{\nu})=(\hat{\mu}_{\hat{\zeta}},\hat{\nu
}_{\hat{\zeta}})\in\mathcal{B^{\prime}}$ as in lemma \ref{hatnu11}.
\end{lemma}

\noindent\textbf{Proof:} We have $\forall(\beta,\gamma)\in\mathcal{B},
\ \forall(\mu, \nu)\in\mathcal{B^{\prime}},\ \forall\zeta\in(0,\infty
),\ \forall x>0$,
\begin{align*}
&  \ \ \ \ E\big[\int_{0}^{T}(\Gamma_{0,s}^{\hat\beta,\hat\gamma}F(s,\hat
\beta_{s},\hat\gamma_{s})+\hat\zeta N^{\hat\mu, \hat\nu}_{0,s}\tilde
b(s,\hat\mu_{s}, \hat\nu_{s})) ds+\Gamma_{0,T}^{\hat\beta,\hat\gamma}\tilde
u\big(\hat\zeta\frac{ N^{\hat\mu, \hat\nu}_{0,T}}{\Gamma_{0,T}^{\hat\beta
,\hat\gamma}}\big)+\hat\zeta x\big]\\
&  =\tilde{V}(\hat{\zeta})+\hat{\zeta}x\\
&  \leq\tilde{V}({\zeta})+\zeta x\\
&  \leq E\big[\int_{0}^{T}(\Gamma_{0,s}^{\beta,\gamma}F(s,\beta_{s},\gamma
_{s})+\zeta N^{\mu, \nu}_{0,s}\tilde b(s,\mu_{s}, \nu_{s})) ds+\Gamma
_{0,T}^{\beta,\gamma}\tilde u\big(\zeta\frac{ N^{\mu, \nu}_{0,T}}{\Gamma
_{0,T}^{\beta,\gamma}}\big)+\zeta x\big].
\end{align*}
This completes the proof. $\ \ \ \ \ \Box$

\begin{assumption}
\label{igrowth} There exist some nonnegative numbers $k_{3},k_{4},p_{1}$ such
that
\[
|I(\zeta)|\leq k_{3}+k_{4}\zeta^{p_{1}},\ \forall\zeta>0.
\]

\end{assumption}

Our main result is the following theorem.

\begin{theorem}
Under Assumption \ref{drift}, \ref{assf}, \ref{assu}, \ref{bound} and
\ref{igrowth}, let $(\hat{\zeta},\hat{\mu},\hat{\nu},\hat{\beta},\hat{\gamma
})$ as in lemma \ref{hatnuzeta1} and define
\[
\hat{\xi}=I\big(\hat{\zeta}\frac{N_{0,T}^{\hat{\mu},\hat{\nu}}}{\Gamma
_{0,T}^{\hat{\beta},\hat{\gamma}}}\big)\ a.s.
\]
Then we have $\forall\xi\in\mathcal{A}(x),\ \forall(\beta,\gamma
)\in\mathcal{B}$,%
\begin{align}
&  \ \ \ \ E\big[\int_{0}^{T}\Gamma_{0,s}^{\hat{\beta},\hat{\gamma}}%
F(s,\hat{\beta}_{s},\hat{\gamma}_{s})ds+\Gamma_{0,T}^{\hat{\beta},\hat{\gamma
}}u(\xi)\big]\nonumber\\
&  \leq E\big[\int_{0}^{T}\Gamma_{0,s}^{\hat{\beta},\hat{\gamma}}%
F(s,\hat{\beta}_{s},\hat{\gamma}_{s})ds+\Gamma_{0,T}^{\hat{\beta},\hat{\gamma
}}u(\hat{\xi})\big]\nonumber\\
&  \leq E\big[\int_{0}^{T}\Gamma_{0,s}^{\beta,\gamma}F(s,\beta_{s},\gamma
_{s})ds+\Gamma_{0,T}^{\beta,\gamma}u(\hat{\xi})\big].
\end{align}
That is to say, $(\hat{\xi},\hat{\beta},\hat{\gamma})$ is a saddle point of
problem (\ref{varformulation}).
\end{theorem}

\noindent\textbf{Proof:} The proof is divided into three steps.

\textbf{Step 1.} We prove that $\forall(\beta,\gamma)\in\mathcal{B}$,
\[
E\big[\int_{0}^{T}\Gamma_{0,s}^{\hat{\beta},\hat{\gamma}}F(s,\hat{\beta}%
_{s},\hat{\gamma}_{s})ds+\Gamma_{0,T}^{\hat{\beta},\hat{\gamma}}u(\hat{\xi
})\big]\leq E\big[\int_{0}^{T}\Gamma_{0,s}^{\beta,\gamma}F(s,\beta_{s}%
,\gamma_{s})ds+\Gamma_{0,T}^{\beta,\gamma}u(\hat{\xi})\big].
\]

By the boundedness of $\hat{\mu},\hat{\nu},\hat{\beta},\hat{\gamma}$, we have
$\frac{N_{0,T}^{\hat{\mu},\hat{\nu}}}{\Gamma_{0,T}^{\hat{\beta},\hat{\gamma}}%
}\in L^{p}$ for any $p\geq1$. Then, Assumption \ref{igrowth} guarantees
$\hat{\xi}\in L^{2}$. By lemma \ref{hatnu11},
\begin{align*}
\tilde{V}(\hat{\zeta})  &  =\inf_{\substack{(\beta,\gamma)\in\mathcal{B}%
\\(\mu,\nu)\in\mathcal{B^{\prime}}}}E\big[\int_{0}^{T}(\Gamma_{0,s}%
^{\beta,\gamma}F(s,\beta_{s},\gamma_{s})+\hat{\zeta}N_{0,s}^{\mu,\nu}\tilde
{b}(s,\mu_{s},\nu_{s}))ds+\Gamma_{0,T}^{\beta,\gamma}\tilde{u}\big(\hat{\zeta
}\frac{N_{0,T}^{\mu,\nu}}{\Gamma_{0,T}^{\beta,\gamma}}\big)\big].\\
&  =E\big[\int_{0}^{T}(\Gamma_{0,s}^{\hat{\beta},\hat{\gamma}}F(s,\hat{\beta
}_{s},\hat{\gamma}_{s})+\hat{\zeta}N_{0,s}^{\hat{\mu},\hat{\nu}}\tilde
{b}(s,\hat{\mu}_{s},\hat{\nu}_{s}))ds+\Gamma_{0,T}^{\hat{\beta},\hat{\gamma}%
}\tilde{u}\big(\hat{\zeta}\frac{N_{0,T}^{\hat{\mu},\hat{\nu}}}{\Gamma
_{0,T}^{\hat{\beta},\hat{\gamma}}}\big)\big].
\end{align*}
It yields that $(\hat{\beta},\hat{\gamma})$ is an optimal control of the
following optimization problem:
\begin{equation}
\inf_{(\beta,\gamma)\in\mathcal{B}}E\big[\int_{0}^{T}(\Gamma_{0,s}%
^{\beta,\gamma}F(s,\beta_{s},\gamma_{s})ds+\Gamma_{0,T}^{\beta,\gamma}%
\tilde{u}\big(\hat{\zeta}\frac{N_{0,T}^{\hat{\mu},\hat{\nu}}}{\Gamma
_{0,T}^{\beta,\gamma}}\big)\big]
\end{equation}
subject to%
\[%
\begin{cases}
d\Gamma_{0,t}^{\beta,\gamma}=\Gamma_{0,t}^{\beta,\gamma}\beta_{t}%
dt+\Gamma_{0,t}^{\beta,\gamma}\gamma_{t}^{\prime}d{W}_{t},\\
\Gamma_{0,0}^{\beta,\gamma}=1,
\end{cases}
\]
where $\Gamma_{0,t}^{\beta,\gamma}$ is the state process at time $t$. Applying
the maximum principle in Peng \cite{Pe}, we obtain a necessary condition for
$(\hat{\beta},\hat{\gamma})$:
\begin{equation}
F(t,\beta_{t},\gamma_{t})+p_{t}\beta_{t}+q_{t}\gamma_{t}\geq F(t,\hat{\beta
}_{t},\hat{\gamma}_{t})+p_{t}\hat{\beta}_{t}+q_{t}\hat{\gamma}_{t}%
,\ \forall(\beta,\gamma)\in\mathcal{B}, \label{qmp}%
\end{equation}
where $(p_{t},q_{t})$ is the solution of the following adjoint equation%
\begin{equation}%
\begin{cases}
-dp_{t}=\big(F(t,\hat{\beta}_{t},\hat{\gamma}_{t})+p_{t}\hat{\beta}_{t}%
+q_{t}^{\prime}\hat{\gamma}_{t}\big)dt-q_{t}^{\prime}d{W}_{t},\\
p_{T}=u\big(I(\hat{\zeta}\frac{N_{0,T}^{\hat{\mu},\hat{\nu}}}{\Gamma
_{0,T}^{\hat{\beta},\hat{\gamma}}})\big).
\end{cases}
\label{adjoint}%
\end{equation}

$\forall(\beta,\gamma)\in\mathcal{B}$, let $(y_{t},z_{t})$ and $(\tilde{y}%
_{t},\tilde{z}_{t})$ be the unique solutions for the following two linear
BSDEs respectively,%

\begin{equation}
y_{t}=u(\hat{\xi})+\int_{t}^{T}\big(y_{s}\hat{\beta}_{s}+z_{s}^{\prime}%
\hat{\gamma}_{s}+F(s,\hat{\beta}_{s},\hat{\gamma}_{s})\big)ds-\int_{t}%
^{T}z_{s}^{\prime}d{W}_{s}, \label{optu}%
\end{equation}%
\begin{equation}
\tilde{y}_{t}=u(\hat{\xi})+\int_{t}^{T}\big(\tilde{y}_{s}\beta_{s}+\tilde
{z}_{s}^{\prime}\gamma_{s}+F(s,\beta_{s},\gamma_{s})\big)ds-\int_{t}^{T}%
\tilde{z}_{s}^{\prime}d{W}_{s}.
\end{equation}

Note that (\ref{adjoint}) and (\ref{optu}) share the same solution. Thus,
(\ref{qmp}) can be written as%
\begin{equation}
F(t,\beta_{t},\gamma_{t})+y_{t}\beta_{t}+z_{t}\gamma_{t}\geq F(t,\hat{\beta
}_{t},\hat{\gamma}_{t})+y_{t}\hat{\beta}_{t}+z_{t}\hat{\gamma}_{t}%
,\ \forall(\beta,\gamma)\in\mathcal{B}. \label{qmp1}%
\end{equation}
By (\ref{qmp1}) and the comparison theorem of BSDE, $y_{t}\leq\tilde{y}%
_{t},\ a.s.$. Especially, we have $y_{0}\leq\tilde{y}_{0}.$ It is easy to see
that
\[
y_{0}=E\big[\int_{0}^{T}\Gamma_{0,s}^{\hat{\beta},\hat{\gamma}}F(s,\hat{\beta
}_{s},\hat{\gamma}_{s})ds+\Gamma_{0,T}^{\hat{\beta},\hat{\gamma}}u(\hat{\xi
})\big],
\]
and
\[
\tilde{y}_{0}=E\big[\int_{0}^{T}\Gamma_{0,s}^{\beta,\gamma}F(s,\beta
_{s},\gamma_{s})ds+\Gamma_{0,T}^{\beta,\gamma}u(\hat{\xi})\big].
\]
Thus,
\begin{equation}
E\big[\int_{0}^{T}\Gamma_{0,s}^{\hat{\beta},\hat{\gamma}}F(s,\hat{\beta}%
_{s},\hat{\gamma}_{s})ds+\Gamma_{0,T}^{\hat{\beta},\hat{\gamma}}u(\hat{\xi
})\big]\leq E\big[\int_{0}^{T}\Gamma_{0,s}^{\beta,\gamma}F(s,\beta_{s}%
,\gamma_{s})ds+\Gamma_{0,T}^{\beta,\gamma}u(\hat{\xi})\big],\ \forall
(\beta,\gamma)\in\mathcal{B}.\nonumber
\end{equation}

\textbf{Step 2.} We prove that
\begin{equation}
E[\hat{\xi}N_{0,T}^{\hat{\mu},\hat{\nu}}]=x+E\int_{0}^{T}N_{0,s}^{\hat{\mu
},\hat{\nu}}\tilde{b}(s,\hat{\mu}_{s},\hat{\nu}_{s})ds. \label{optcon11}%
\end{equation}

From the convexity of $\tilde{u}$ and Lemma \ref{hatnu11}, we get $\tilde
{V}(\cdot)$ is convex. Fix $\zeta>0$. Then for any $\delta>0$, we have
\begin{align*}
\frac{\tilde{V}(\zeta+\delta)-\tilde{V}(\zeta)}{\delta}  &  \leq
\frac{E\big[\Gamma_{0,T}^{\hat{\beta},\hat{\gamma}}\tilde{u}\big((\zeta
+\delta)\frac{N_{0,T}^{\hat{\mu},\hat{\nu}}}{\Gamma_{0,T}^{\hat{\beta}%
,\hat{\gamma}}}\big)-\Gamma_{0,T}^{\hat{\beta},\hat{\gamma}}\tilde
{u}\big(\zeta\frac{N_{0,T}^{\hat{\mu},\hat{\nu}}}{\Gamma_{0,T}^{\hat{\beta
},\hat{\gamma}}}\big)\big]}{\delta}+E\int_{0}^{T}N_{0,s}^{\hat{\mu},\hat{\nu}%
}\tilde{b}(s,\hat{\mu}_{s},\hat{\nu}_{s})ds\\
&  \leq E\big[N_{0,T}^{\hat{\mu},\hat{\nu}}\tilde{u}^{\prime}\big((\zeta
+\delta)\frac{N_{0,T}^{\hat{\mu},\hat{\nu}}}{\Gamma_{0,T}^{\hat{\beta}%
,\hat{\gamma}}}\big)\big]+E\int_{0}^{T}N_{0,s}^{\hat{\mu},\hat{\nu}}\tilde
{b}(s,\hat{\mu}_{s},\hat{\nu}_{s})ds\\
&  =-E\big[N_{0,T}^{\hat{\mu},\hat{\nu}}I\big((\zeta+\delta)\frac
{N_{0,T}^{\hat{\mu},\hat{\nu}}}{\Gamma_{0,T}^{\hat{\beta},\hat{\gamma}}%
}\big)\big]+E\int_{0}^{T}N_{0,s}^{\hat{\mu},\hat{\nu}}\tilde{b}(s,\hat{\mu
}_{s},\hat{\nu}_{s})ds.
\end{align*}
By Levi's lemma,
\begin{equation}
\lim_{\delta\rightarrow0+}\frac{\tilde{V}(\zeta+\delta)-\tilde{V}(\zeta
)}{\delta}\leq-E\big[N_{0,T}^{\hat{\mu},\hat{\nu}}I\big(\zeta\frac
{N_{0,T}^{\hat{\mu},\hat{\nu}}}{\Gamma_{0,T}^{\hat{\beta},\hat{\gamma}}%
}\big)\big]+E\int_{0}^{T}N_{0,s}^{\hat{\mu},\hat{\nu}}\tilde{b}(s,\hat{\mu
}_{s},\hat{\nu}_{s})ds.
\end{equation}
We can similarly deduce that
\begin{equation}
\lim_{\delta\rightarrow0+}\frac{\tilde{V}(\zeta)-\tilde{V}(\zeta-\delta
)}{\delta}\geq-E\big[N_{0,T}^{\hat{\mu},\hat{\nu}}I\big(\zeta\frac
{N_{0,T}^{\hat{\mu},\hat{\nu}}}{\Gamma_{0,T}^{\hat{\beta},\hat{\gamma}}%
}\big)\big]+E\int_{0}^{T}N_{0,s}^{\hat{\mu},\hat{\nu}}\tilde{b}(s,\hat{\mu
}_{s},\hat{\nu}_{s})ds.
\end{equation}
So $\tilde{V}(\cdot)$ is differentiable on $(0,\infty)$ and
\begin{equation}
\tilde{V}^{\prime}(\zeta)=-E\big[N_{0,T}^{\hat{\mu},\hat{\nu}}I\big(\zeta
\frac{N_{0,T}^{\hat{\mu},\hat{\nu}}}{\Gamma_{0,T}^{\hat{\beta},\hat{\gamma}}%
}\big)\big]+E\int_{0}^{T}N_{0,s}^{\hat{\mu},\hat{\nu}}\tilde{b}(s,\hat{\mu
}_{s},\hat{\nu}_{s})ds. \label{diffe}%
\end{equation}
From lemma \ref{hatzeta1}, we know $\hat{\zeta}\in(0,\infty)$ attains the
infimum of $\inf_{\zeta>0}(\tilde{V}(\zeta)+\zeta x)$. Then $\tilde{V}%
^{\prime}(\hat{\zeta})=-x$. Combined with (\ref{diffe}), we derive
(\ref{optcon11}).

\textbf{Step 3.} We prove that $\forall\xi\in\mathcal{A}(x),$%
\[
E\big[\int_{0}^{T}\Gamma_{0,s}^{\hat{\beta},\hat{\gamma}}F(s,\hat{\beta}%
_{s},\hat{\gamma}_{s})ds+\Gamma_{0,T}^{\hat{\beta},\hat{\gamma}}%
u(\xi)\big]\leq E\big[\int_{0}^{T}\Gamma_{0,s}^{\hat{\beta},\hat{\gamma}%
}F(s,\hat{\beta}_{s},\hat{\gamma}_{s})ds+\Gamma_{0,T}^{\hat{\beta},\hat
{\gamma}}u(\hat{\xi})\big].
\]

By the definition of $\tilde{u}(\cdot)$ and (\ref{constraint-xi}), $\forall
\xi\in\mathcal{A}(x),$
\begin{align*}
&  \ \ \ \ E\big[\int_{0}^{T}\Gamma_{0,s}^{\hat{\beta},\hat{\gamma}}%
F(s,\hat{\beta}_{s},\hat{\gamma}_{s})ds+\Gamma_{0,T}^{\hat{\beta},\hat{\gamma
}}u(\xi)\big]\\
&  \leq E\big[\int_{0}^{T}\Gamma_{0,s}^{\hat{\beta},\hat{\gamma}}%
F(s,\hat{\beta}_{s},\hat{\gamma}_{s})ds+\Gamma_{0,T}^{\hat{\beta},\hat{\gamma
}}\tilde{u}\big(\zeta\frac{N_{0,T}^{\hat{\mu},\hat{\nu}}}{\Gamma_{0,T}%
^{\hat{\beta},\hat{\gamma}}}\big)+\zeta\xi N_{0,T}^{\hat{\mu},\hat{\nu}%
}\big]\\
&  \leq E\big[\int_{0}^{T}(\Gamma_{0,s}^{\hat{\beta},\hat{\gamma}}%
F(s,\hat{\beta}_{s},\hat{\gamma}_{s})+\hat{\zeta}N_{0,s}^{\hat{\mu},\hat{\nu}%
}\tilde{b}(s,\hat{\mu}_{s},\hat{\nu}_{s}))ds+\Gamma_{0,T}^{\hat{\beta}%
,\hat{\gamma}}\tilde{u}\big(\hat{\zeta}\frac{N_{0,T}^{\hat{\mu},\hat{\nu}}%
}{\Gamma_{0,T}^{\hat{\beta},\hat{\gamma}}}\big)\big]+\hat{\zeta}x\\
&  =E\big[\int_{0}^{T}\Gamma_{0,s}^{\hat{\beta},\hat{\gamma}}F(s,\hat{\beta
}_{s},\hat{\gamma}_{s})ds+\Gamma_{0,T}^{\hat{\beta},\hat{\gamma}}u(\hat{\xi
})\big].
\end{align*}
The last equality is due to (\ref{optcon11}). \ \ \ \ \ \ \ \ \ \

This completes the proof. $\ \ \ \ \ \Box$

\section{Investors with ambiguity aversion}

For any $d$-dimensional vector $K=(K_{1},...,K_{d})^{\prime}$, denote by $|K|$
the $d$-dimensional vector with $i$th component $|K_{i}|,\ i=1,...,d$; denote
by $\parallel K\parallel$ the Euclidean norm $(%
%TCIMACRO{\dsum \limits_{i=1}^{d}}%
%BeginExpansion
{\displaystyle\sum\limits_{i=1}^{d}}
%EndExpansion
K_{i}^{2})^{\frac{1}{2}}$. We say that two $d$-dimensional vectors
$K\geq\tilde{K}$ if $K_{i}\geq\tilde{K}_{i},\ i=1,...,d$. For any $K,\tilde
{K},\hat{K}\in\mathbb{R}^{d}$, denote by $K\vee\tilde{K}\wedge\hat{K}$ the
$d$-dimensional vector with $i$th component $K_{i}\vee\tilde{K}_{i}\wedge
\hat{K}_{i}$.

We model the utility process (\ref{BSDE}) via
\begin{equation}
f(t,y,z)=-K^{\prime}|z|,
\end{equation}
where $K$ is a given $d$-dimensional vector whose component $K_{i}%
\geq0,\ i=1,2,...,d$. Chen and Epstein \cite{CE} interpreted the term
$-K^{\prime}|z|$ as modeling ambiguity aversion rather than risk aversion.
This special formulation in \cite{CE} is called K-ignorance. When
$K_{i}=0,\ i=1,2,...,d$, it degenerates to the classical expected utility
maximization problem studied in \cite{CC}, \cite{KLS} etc.

For the K-ignorance case, we have
\[
F(\omega,t,\beta,\gamma)\equiv0
\]
and
\[
\mathcal{B}=\{(\beta,\gamma)\mid\beta_{t}=0,\ \gamma_{t}^{\prime}=(\gamma
_{1t},...,\gamma_{dt})\ \text{is}\ \{\mathcal{F}_{t}\}_{t\geq0}%
\text{-progressively measurable and}\ |\gamma_{t}|\leq K,\text{ }%
t\in\lbrack0,T], \ a.s.\}.
\]
Define
\[
\mathcal{B}_{2}=\{\gamma\mid\gamma_{t}^{\prime}=(\gamma_{1t},...,\gamma
_{dt})\ \text{is}\ \{\mathcal{F}_{t}\}_{t\geq0}\text{-progressively measurable
and}\ |\gamma_{t}|\leq K,t\in\lbrack0,T],\ a.s.\}.
\]
For a given $\gamma\in\mathcal{B}_{2}$, define%
\[
\mathcal{B}_{2,t}(\gamma)=\{\tilde{\gamma}\in\mathcal{B}_{2}\mid\tilde{\gamma
}_{s}\equiv\gamma_{s}\text{ on }[0,t],\ a.s.\}.
\]

For some $0<\alpha<1$, set
\[
u(x)=\frac{1}{\alpha}x^{\alpha},\ x\geq0.
\]
It is easy to check that $u$ satisfies Assumption \ref{assu}, \ref{igrowth}
and for any $\zeta>0$,
\begin{align*}
I(\zeta)  &  =\zeta^{\frac{1}{\alpha-1}},\\
\tilde{u}(\zeta)  &  =u(I(\zeta))-\zeta I(\zeta)=\frac{1-\alpha}{\alpha}%
\zeta^{\frac{\alpha}{\alpha-1}}.
\end{align*}

In this section, we assume that the investors have the same recursive utility
as above. In the following, we investigate three different kinds of wealth equations.

\subsection{Linear wealth equation}

In Example \ref{exam-1}, suppose that $r_{t}\equiv0$ and $\sigma_{t}\equiv
I_{d\times d}$. Then, the wealth equation becomes
\[%
\begin{cases}
dX_{t}=\pi_{t}^{\prime}b_{t}dt+\pi_{t}dW_{t},\\
X_{0}=x.
\end{cases}
\]
where $b_{t}$ is a uniformly bounded progressively measurable process.

In this case, $\tilde{b}\equiv0$ and
\[
\mathcal{B}^{\prime}=\{(\mu,\nu)\mid\mu_{t}=0,\ \nu_{t}=b_{t},\text{ }t\in\lbrack0,T], \text{
}a.s.\}.
\]

Then the value function $\tilde{V}(\zeta)$ in (\ref{dua2}) becomes
\begin{equation}
\tilde{V}(\zeta)=\inf_{\gamma\in\mathcal{B}_{2}}E\big[\Gamma_{0,T}^{0,\gamma
}\tilde{u}\big(\zeta\frac{N_{0,T}^{0,b}}{\Gamma_{0,T}^{0,\gamma}%
}\big)\big]=\frac{1-\alpha}{\alpha}\zeta^{\frac{\alpha}{\alpha-1}}\inf
_{\gamma\in\mathcal{B}_{2}}E\big[(N_{0,T}^{0,b})^{\frac{\alpha}{\alpha-1}%
}(\Gamma_{0,T}^{0,\gamma})^{\frac{1}{1-\alpha}}\big]. \label{tildeV1}%
\end{equation}
Define
\[
\tilde{V}(t,\gamma,\zeta)=\frac{1-\alpha}{\alpha}\zeta^{\frac{\alpha}%
{\alpha-1}}\underset{\tilde{\gamma}\in\mathcal{B}_{2,t}(\gamma)}{\text{essinf}%
}E\big[(N_{0,T}^{0,b})^{\frac{\alpha}{\alpha-1}}(\Gamma_{0,T}^{0,\tilde
{\gamma}})^{\frac{1}{1-\alpha}}\big|\mathcal{F}_{t}\big].
\]
We conjecture that $\tilde{V}(t,\gamma,\zeta)$ has the following form
\[
\tilde{V}(t,\gamma,\zeta)=\frac{1-\alpha}{\alpha}\zeta^{\frac{\alpha}%
{\alpha-1}}(N_{0,t}^{0,b})^{\frac{\alpha}{\alpha-1}}(\Gamma_{0,t}^{0,\gamma
})^{\frac{1}{1-\alpha}}e^{\tilde{Y}_{t}},
\]
where $(\tilde{Y},\tilde{Z})$ is the solution of the following BSDE
\begin{equation}
\tilde{Y}_{t}=\int_{t}^{T}g(s,\tilde{Z}_{s})ds-\int_{t}^{T}\tilde{Z}%
_{s}^{\prime}dW_{s}. \label{BSDEqua}%
\end{equation}

The generator $g(t,z)$ of (\ref{BSDEqua}) can be determined via the following
martingale principle in \cite{El}. The readers may also refer to Hu et al
\cite{HIM}.

\begin{lemma}
\label{martingale} The process $\tilde{V}(t,\gamma,\zeta)$ is a submartingale
for any $\gamma\in\mathcal{B}_{2}$ and $\hat{\gamma}$ is the solution of
problem (\ref{tildeV1}) if and only if $\tilde{V}(t,\hat{\gamma},\zeta)$ is a martingale.
\end{lemma}

Applying It\^{o}'s formula to $(N_{0,t}^{0,b})^{\frac{\alpha}{\alpha-1}%
}(\Gamma_{0,t}^{0,\gamma})^{\frac{1}{1-\alpha}}e^{\tilde{Y}_{t}}$,
\begin{align*}
&  \ \ \ \ d(N_{0,t}^{0,b})^{\frac{\alpha}{\alpha-1}}(\Gamma_{0,t}^{0,\gamma
})^{\frac{1}{1-\alpha}}e^{\tilde{Y}_{t}}\\
&  =(N_{0,t}^{0,b})^{\frac{\alpha}{\alpha-1}}(\Gamma_{0,t}^{0,\gamma}%
)^{\frac{1}{1-\alpha}}e^{\tilde{Y}_{t}}\big[-g(t,\tilde{Z}_{t})+\frac{1}%
{2}\parallel\tilde{Z}_{t}\parallel^{2}+\frac{1}{2}\frac{\alpha}{(1-\alpha
)^{2}}\parallel b_{t}+\gamma_{t}\parallel^{2}+\frac{1}{1-\alpha}\tilde{Z}%
_{t}^{\prime}\gamma_{t}+\frac{\alpha}{1-\alpha}\tilde{Z}_{t}^{\prime}%
b_{t}\big]dt\\
&  \ \ \ \ +(N_{0,t}^{0,b})^{\frac{\alpha}{\alpha-1}}(\Gamma_{0,t}^{0,\gamma
})^{\frac{1}{1-\alpha}}e^{\tilde{Y}_{t}}\big[\frac{1}{1-\alpha}\gamma
_{t}^{\prime}+\frac{\alpha}{1-\alpha}b_{t}^{\prime}+\tilde{Z}_{t}^{\prime
}\big]dW_{t}.
\end{align*}
According to Lemma \ref{martingale}, we have
\[
-g(t,\tilde{Z}_{t})+\frac{1}{2}\parallel\tilde{Z}_{t}\parallel^{2}+\frac{1}%
{2}\frac{\alpha}{(1-\alpha)^{2}}\parallel b_{t}+\gamma_{t}\parallel^{2}%
+\frac{1}{1-\alpha}\gamma_{t}^{\prime}\tilde{Z}_{t}+\frac{\alpha}{1-\alpha
}b_{t}^{\prime}\tilde{Z}_{t}\geq0,\ \forall\gamma\in\mathcal{B}_{2}.
\]

For $z\in\mathbb{R}^{d}$ and $t\in\lbrack0,T]$, define
\begin{align*}
g(t,z)  &  =\underset{\gamma\in\mathcal{B}_{2}}{\text{essinf}}\big[\frac{1}%
{2}\frac{\alpha}{(1-\alpha)^{2}}\parallel b_{t}+\gamma_{t}\parallel^{2}%
+\frac{1}{1-\alpha}\gamma_{t}^{\prime}z+\frac{\alpha}{1-\alpha}b_{t}^{\prime
}z+\frac{1}{2}\parallel z\parallel^{2}\big]\\
&  =\frac{1}{2}\frac{\alpha}{(1-\alpha)^{2}}\underset{\gamma\in\mathcal{B}%
_{2}}{\text{essinf}}\big[\parallel\gamma_{t}+b_{t}+\frac{1-\alpha}{\alpha
}z\parallel^{2}-2\frac{(1-\alpha)^{2}}{\alpha}b_{t}^{\prime}z-\frac
{(1-\alpha)^{3}}{\alpha^{2}}\parallel z\parallel^{2}\big]\\
&  =\frac{1}{2}\frac{\alpha}{(1-\alpha)^{2}}\big[\text{dist}_{\mathcal{B}_{2}%
}^{2}(b_{t}+\frac{1-\alpha}{\alpha}z)-2\frac{(1-\alpha)^{2}}{\alpha}%
b_{t}^{\prime}z-\frac{(1-\alpha)^{3}}{\alpha^{2}}\parallel z\parallel
^{2}\big],
\end{align*}
where $\text{dist}_{\mathcal{B}_{2}}^{2}(b_{t}+\frac{1-\alpha}{\alpha
}z)=\underset{\gamma\in\mathcal{B}_{2}}{\text{essinf}}\parallel\gamma
_{t}+b_{t}+\frac{1-\alpha}{\alpha}z\parallel^{2}$. By the result of Kobylanski
\cite{Ko}, the quadratic BSDE (\ref{BSDEqua}) has a unique solution
$(\tilde{Y},\tilde{Z})$.

Then the infimum in (\ref{tildeV1}) is attained at
\[
\hat{\gamma}_{t}=(-K)\vee(-b_{t}-\frac{1-\alpha}{\alpha}\tilde{Z}_{t})\wedge
K,\text{ }
t\in\lbrack0,T],\ a.s.\]
and
\[
\tilde{V}(\zeta)=\frac{1-\alpha}{\alpha}\zeta^{\frac{\alpha}{\alpha-1}%
}E\big[(N_{0,T}^{0,b})^{\frac{\alpha}{\alpha-1}}(\Gamma_{0,T}^{0,\hat{\gamma}%
})^{\frac{1}{1-\alpha}}\big],\ \zeta>0.
\]
The second infimum in (\ref{dua11}) is attained at
\[
\hat{\zeta}=x^{\alpha-1}\big(E[(N_{0,T}^{0,b})^{\frac{\alpha}{\alpha-1}%
}(\Gamma_{0,T}^{0,\hat{\gamma}})^{\frac{1}{1-\alpha}}]\big)^{1-\alpha}.
\]
Thus, the optimal terminal wealth is given by
\[
\hat{\xi}=I(\hat{\zeta}\frac{N_{0,T}^{0,b}}{\Gamma_{0,T}^{0,\hat{\gamma}}%
})=x\big(E[(N_{0,T}^{0,b})^{\frac{\alpha}{\alpha-1}}(\Gamma_{0,T}%
^{0,\hat{\gamma}})^{\frac{1}{1-\alpha}}]\big)^{-1}(N_{0,T}^{0,b})^{\frac
{1}{\alpha-1}}(\Gamma_{0,T}^{0,\hat{\gamma}})^{\frac{1}{1-\alpha}}.
\]
It is easy to check the following propositions.

\begin{proposition}
When $b_{t}$ is a deterministic function, we have $\tilde{Z}_{t}=0$ and
\[
\hat{\gamma}_{t}=(-K)\vee(-b_{t})\wedge K,\ t\in\lbrack0,T].
\]

\end{proposition}

\subsection{Higher interest rate for borrowing}

In this subsection, for simplicity, we assume that all variables are $1$-dimentional.

In Example \ref{exhigher}, the investor is allowed to borrow money with
interest rate $R_{t}\geq r_{t},\ t\in\lbrack0,T]$. Suppose that $b,$ $R,$ $r$
are deterministic continuous functions and $\sigma_{t}\equiv1$. Then the
wealth equation becomes
\[%
\begin{cases}
dX_{t}=\big(r_{t}X_{t}+\pi_{t}(b_{t}-r_{t})-(R_{t}-r_{t})(X_{t}-\pi_{t}%
)^{-}\big)dt+\pi_{t}dW_{t},\\
X_{0}=x.
\end{cases}
\]

In this case, $\tilde{b}\equiv0$,
\[
\mathcal{B^{\prime}}=\{(\mu,\nu)\mid(\mu_{t},\nu_{t})\ \text{is}%
\ \{\mathcal{F}_{t}\}_{t\geq0}\text{-progressively measurable and }r_{t}%
\leq\mu_{t}\leq R_{t},\text{ }\mu_{t}+\nu_{t}=b_{t},\text{ }t\in\lbrack0,T], \ a.s.\}.
\]
and%
\[
\mathcal{B}_{2}^{\prime}=\{\mu\mid\mu_{t}\ \text{is}\ \{\mathcal{F}%
_{t}\}_{t\geq0}\text{-progressively measurable and}\ r_{t}\leq\mu_{t}\leq
R_{t},\text{ }t\in\lbrack0,T],\ a.s. \}.
\]

The value function $\tilde{V}(\zeta)$ in (\ref{dua2}) becomes
\begin{align}
\tilde{V}(\zeta)  &  =\inf_{\substack{(\beta,\gamma)\in\mathcal{B}\\(\mu
,\nu)\in\mathcal{B^{\prime}}}}E\big[\Gamma_{0,T}^{\beta,\gamma}\tilde
{u}\big(\zeta\frac{N_{0,T}^{\mu,\nu}}{\Gamma_{0,T}^{\beta,\gamma}%
}\big)\big]\ \nonumber\label{tildeV3}\\
&  =\inf_{\mu\in\mathcal{B}_{2}^{\prime},\gamma\in\mathcal{B}_{2}}%
E\big[\Gamma_{0,T}^{0,\gamma}\tilde{u}\big(\zeta\frac{N_{0,T}^{\mu,b-\mu}%
}{\Gamma_{0,T}^{0,\gamma}}\big)\big]\ \nonumber\\
&  =\frac{1-\alpha}{\alpha}\zeta^{\frac{\alpha}{\alpha-1}}\inf_{\mu
\in\mathcal{B}_{2}^{\prime},\gamma\in\mathcal{B}_{2}}E\big[(N_{0,T}^{\mu
,b-\mu})^{\frac{\alpha}{\alpha-1}}(\Gamma_{0,T}^{0,\gamma})^{\frac{1}%
{1-\alpha}}\big].
\end{align}

Define
\[
\tilde{V}(t,\mu,\gamma,\zeta)=\frac{1-\alpha}{\alpha}\zeta^{\frac{\alpha
}{\alpha-1}}\underset{\tilde{\mu}\in\mathcal{B}_{2,t}^{\prime}(\mu
),\tilde{\gamma}\in\mathcal{B}_{2,t}(\gamma)}{\text{essinf}}E\big[(N_{0,T}%
^{\mu,b-\tilde{\mu}})^{\frac{\alpha}{\alpha-1}}(\Gamma_{0,T}^{0,\tilde{\gamma
}})^{\frac{1}{1-\alpha}}\big|\mathcal{F}_{t}\big],
\]
where $\mathcal{B}_{2,t}^{\prime}(\mu)=\{\tilde{\mu}\in\mathcal{B}_{2}%
^{\prime}\mid\tilde{\mu}_{s}=\mu_{s}\text{ on }[0,t].\}$.

We conjecture that $\tilde{V}(t,\mu,\gamma,\zeta)$ has the following form:
\[
\tilde{V}(t,\mu,\gamma,\zeta)=\frac{1-\alpha}{\alpha}\zeta^{\frac{\alpha
}{\alpha-1}}(N_{0,t}^{\mu,b-\mu})^{\frac{\alpha}{\alpha-1}}(\Gamma
_{0,t}^{0,\gamma})^{\frac{1}{1-\alpha}}e^{\tilde{Y}_{t}},
\]
where $(\tilde{Y},\tilde{Z})$ is the solution of the BSDE
\begin{equation}
\tilde{Y}_{t}=\int_{t}^{T}g(s,\tilde{Z}_{s})ds-\int_{t}^{T}\tilde{Z}_{s}%
dW_{s}. \label{BSDEqua3}%
\end{equation}

Similarly, $\forall\mu\in\mathcal{B}_{2}^{\prime}$ and $\gamma\in
\mathcal{B}_{2}$, the process $\tilde{V}(t,\mu,\gamma,\zeta)$ is a
submartingale. $\hat{\mu}$ and $\hat{\gamma}$ are the optimal solutions of
problem (\ref{tildeV3}) if and only if $\tilde{V}(t,\hat{\mu},\hat{\gamma
},\zeta)$ is a martingale. Now we determine $g(t,z)$ via this martingale principle.

Applying It\^{o}'s formula to $(N_{0,t}^{\mu,b-\mu})^{\frac{\alpha}{\alpha-1}%
}(\Gamma_{0,t}^{0,\gamma})^{\frac{1}{1-\alpha}}e^{\tilde{Y}_{t}}$,
\begin{align*}
&  \ \ \ \ d(N_{0,t}^{\mu,b-\mu})^{\frac{\alpha}{\alpha-1}}(\Gamma
_{0,t}^{0,\gamma})^{\frac{1}{1-\alpha}}e^{\tilde{Y}_{t}}\\
&  =(N_{0,t}^{\mu,b-\mu})^{\frac{\alpha}{\alpha-1}}(\Gamma_{0,t}^{0,\gamma
})^{\frac{1}{1-\alpha}}e^{\tilde{Y}_{t}}\big[-g(t,\tilde{Z}_{t})+\frac{1}%
{2}\tilde{Z}_{t}^{2}+\frac{1}{2}\frac{\alpha}{(1-\alpha)^{2}}(b_{t}+\gamma
_{t}-\mu_{t})^{2}+\frac{\alpha}{1-\alpha}\mu_t\\
&  \ \ +\frac{1}{1-\alpha}\tilde{Z}_{t}(\gamma_{t}+\alpha(b_{t}-\mu
_{t}))\big]dt+(N_{0,t}^{\mu,b-\mu})^{\frac{\alpha}{\alpha-1}}(\Gamma
_{0,t}^{0,\gamma})^{\frac{1}{1-\alpha}}e^{\tilde{Y}_{t}}\big[\frac{1}%
{1-\alpha}\gamma_{t}+\frac{\alpha}{1-\alpha}(b_{t}-\mu_{t})+\tilde{Z}%
_{t}\big]dW_{t}.
\end{align*}

Define
\[
g(t,z)=\underset{\mu\in\mathcal{B}_{2}^{\prime},\gamma\in\mathcal{B}%
_{2}}{\text{essinf}}\big[\frac{1}{2}\frac{\alpha}{(1-\alpha)^{2}}(b_{t}%
+\gamma_{t}-\mu_{t})^{2}+\frac{\alpha}{1-\alpha}\mu_{t}+\frac{1}{1-\alpha
}z(\gamma_{t}+\alpha(b_{t}-\mu_{t}))+\frac{1}{2}z^{2}\big].
\]
Since $b_{t}$ is a deterministic function, $g(t,z)$ is also deterministic. By
the existence and uniqueness theorem of BSDE, the solution of (\ref{BSDEqua3})
satisfies $\tilde{Z}_{t}=0,$ $t\in\lbrack0,T]$. Thus the optimal solutions
$\hat{\mu}$ and $\hat{\gamma}$ attain the infimum in
\begin{equation}
g(t,0)=\frac{1}{2}\frac{\alpha}{(1-\alpha)^{2}}\underset{\mu\in\mathcal{B}%
_{2}^{\prime},\gamma\in\mathcal{B}_{2}}{\text{inf}}[(b_{t}+\gamma_{t}%
-\mu_{t})^{2}+2(1-\alpha)\mu_{t}]. \label{ghigher}%
\end{equation}

Set
\[
H(\mu,\gamma)=(b_{t}+\gamma-\mu)^{2}+2(1-\alpha)\mu.
\]
It is easy to check that the following equations
\[%
\begin{cases}
\frac{\partial H}{\partial\mu}=-2(b_{t}+\gamma-\mu_{t})+2(1-\alpha)=0,\\
\frac{\partial H}{\partial\gamma}=2(b_{t}+\gamma-\mu)=0
\end{cases}
\]
have no solutions. So the infimum in (\ref{ghigher}) must attained at the
boundary of the region $[r_{t},R_{t}]\times\lbrack-K,K]$.

For the optimal solutions $\hat{\mu}$ and $\hat{\gamma}$, we have
\[
\tilde{V}(\zeta)=\frac{1-\alpha}{\alpha}\zeta^{\frac{\alpha}{\alpha-1}%
}E\big[(N_{0,T}^{\hat{\mu},b-\hat{\mu}})^{\frac{\alpha}{\alpha-1}}%
(\Gamma_{0,T}^{0,\hat{\gamma}})^{\frac{1}{1-\alpha}}\big],\ \zeta>0.
\]
The second infimum in (\ref{dua11}) is attained at
\[
\hat{\zeta}=x^{\alpha-1}\big(E[(N_{0,T}^{\hat{\mu},b-\hat{\mu}})^{\frac
{\alpha}{\alpha-1}}(\Gamma_{0,T}^{0,\hat{\gamma}})^{\frac{1}{1-\alpha}%
}]\big)^{1-\alpha}.
\]
Thus, the optimal terminal wealth is given by
\[
\hat{\xi}=I(\hat{\zeta}\frac{N_{0,T}^{\hat{\mu},b-\hat{\mu}}}{\Gamma
_{0,T}^{0,\hat{\gamma}}})=x\big(E[(N_{0,T}^{\hat{\mu},b-\hat{\mu}}%
)^{\frac{\alpha}{\alpha-1}}(\Gamma_{0,T}^{0,\hat{\gamma}})^{\frac{1}{1-\alpha
}}]\big)^{-1}(N_{0,T}^{\hat{\mu},b-\hat{\mu}})^{\frac{1}{\alpha-1}}%
(\Gamma_{0,T}^{0,\hat{\gamma}})^{\frac{1}{1-\alpha}}.
\]

We can easily deduce the following propositions.

\begin{proposition}
When $K\equiv0$, we obtain $\hat{\gamma}\equiv0$ and $\hat{\mu}_{t}=r_{t}%
\vee(b_{t}-1+\alpha)\wedge R_{t}$. This coincides with the result of Appendix
B in Cvitanic and Karatzas \cite{CK}.
\end{proposition}

\begin{proposition}
If $\frac{1-\alpha}{2}\leq K\leq b_{t}-R_{t},\ t\in\lbrack0,T]$, then the
infimum in (\ref{ghigher}) attained at $\hat{\gamma}_{t}=-K$, and $\hat{\mu
}_{t}=r_{t}\vee(b_{t}-K-1+\alpha)\wedge R_{t},\ t\in\lbrack0,T]$.
\end{proposition}

\begin{remark}
When $b,R,r$ are bounded progressively measurable processes, $g(t,z)$ is no
longer deterministic. Similar analysis as in Theorem 7 of Hu et al \cite{HIM},
we can prove the quadratic BSDE (\ref{BSDEqua3}) has a unique solution
$(\tilde{Y}_{t},\tilde{Z}_{t})$. Thanks to the boundness, closeness and convexity of
$\mathcal{B}_{2}$ and $\mathcal{B}_{2}^{\prime}$, there exists a pair
$(\hat{\mu},\hat{\gamma})$ which attains the infimium of $g(t,\tilde{Z}_{t})$.
\end{remark}

\subsection{Large investor}

Suppose that $r\equiv0,\ \sigma\equiv I_{d\times d}$ and $b_{t}$ is a
deterministic continuous bounded function in Example \ref{exam-2}. Then, the
wealth equation becomes%
\[%
\begin{cases}
dX_{t}=\big(b_{t}^{\prime}\pi_{t}-\varepsilon^{\prime}|\pi_{t}|\big)dt+\pi
_{t}^{\prime}\sigma_{t}dW_{t},\\
X_{0}=x.
\end{cases}
\]

In this case, $\tilde{b}\equiv0$,
\[
\mathcal{B}^{\prime}=\{(\mu,\nu)\mid\mu_{t}=0,\ \nu_{t}=b_{t}-\delta
_{t},\ \delta^{\prime}=(\delta_{1},...,\delta_{d})\ \text{is}\ \{\mathcal{F}%
_{t}\}_{t\geq0}\text{-progressively measurable and}\ |\delta_{t}%
|\leq\varepsilon,\text{ }t\in\lbrack0,T], \ a.s.\}.
\]
and%
\[
\mathcal{B}_{2}^{\prime}=\{\delta\mid\delta_{t}^{\prime}=(\delta
_{1t},...,\delta_{dt})\ \text{is}\ \{\mathcal{F}_{t}\}_{t\geq0}%
\text{-progressively measurable and}\ |\delta_{t}|\leq\varepsilon,\text{ }%
t\in\lbrack0,T], \ a.s. \}.
\]

The value function $\tilde{V}(\zeta)$ in (\ref{dua2}) becomes
\begin{align}
\tilde{V}(\zeta) &  =\inf_{\substack{(\beta,\gamma)\in\mathcal{B}\\(\mu
,\nu)\in\mathcal{B^{\prime}}}}E\big[\Gamma_{0,T}^{\beta,\gamma}\tilde
{u}\big(\zeta\frac{N_{0,T}^{\mu,\nu}}{\Gamma_{0,T}^{\beta,\gamma}%
}\big)\big]\ \nonumber\label{tildeV2}\\
&  =\inf_{\delta\in\mathcal{B}_{2}^{\prime},\gamma\in\mathcal{B}_{2}%
}E\big[\Gamma_{0,T}^{0,\gamma}\tilde{u}\big(\zeta\frac{N_{0,T}^{0,b-\delta}%
}{\Gamma_{0,T}^{0,\gamma}}\big)\big]\ \nonumber\\
&  =\frac{1-\alpha}{\alpha}\zeta^{\frac{\alpha}{\alpha-1}}\inf_{\delta
\in\mathcal{B}_{2}^{\prime},\gamma\in\mathcal{B}_{2}}E\big[(N_{0,T}%
^{0,b-\delta})^{\frac{\alpha}{\alpha-1}}(\Gamma_{0,T}^{0,\gamma})^{\frac
{1}{1-\alpha}}\big].
\end{align}

Consider
\[
\tilde{V}(t,\delta,\gamma,\zeta):=\frac{1-\alpha}{\alpha}\zeta^{\frac{\alpha
}{\alpha-1}}\underset{\tilde{\delta}\in\mathcal{B}_{2,t}^{\prime}%
(\delta),\tilde{\gamma}\in\mathcal{B}_{2,t}(\gamma)}{\text{essinf}%
}E\big[(N_{0,T}^{0,b-\tilde{\delta}})^{\frac{\alpha}{\alpha-1}}(\Gamma
_{0,T}^{0,\tilde{\gamma}})^{\frac{1}{1-\alpha}}\big|\mathcal{F}_{t}\big],
\]
where $\mathcal{B}_{2,t}^{\prime}(\delta)=\{\tilde{\delta}\in\mathcal{B}%
_{2}^{\prime}\mid\tilde{\delta}_{s}=\delta_{s}\text{ on }[0,t]\}$.

We conjecture that $\tilde{V}(t,\delta,\gamma,\zeta)$ has the following form:
\[
\tilde{V}(t,\delta,\gamma,\zeta)=\frac{1-\alpha}{\alpha}\zeta^{\frac{\alpha
}{\alpha-1}}(N_{0,t}^{0,b-\delta})^{\frac{\alpha}{\alpha-1}}(\Gamma
_{0,t}^{0,\gamma})^{\frac{1}{1-\alpha}}e^{\tilde{Y}_{t}},
\]
where $(\tilde{Y},\tilde{Z})$ is the solution of the BSDE
\begin{equation}
\tilde{Y}_{t}=\int_{t}^{T}g(s,\tilde{Z}_{s})ds-\int_{t}^{T}\tilde{Z}%
_{s}^{\prime}dW_{s}.\label{BSDEqua2}%
\end{equation}

Applying It\^{o}'s formula to $(N_{0,t}^{0,b-\delta})^{\frac{\alpha}{\alpha
-1}}(\Gamma_{0,t}^{0,\gamma})^{\frac{1}{1-\alpha}}e^{\tilde{Y}_{t}}$,
\begin{align*}
&  \ \ \ \ d(N_{0,t}^{0,b-\delta})^{\frac{\alpha}{\alpha-1}}(\Gamma
_{0,t}^{0,\gamma})^{\frac{1}{1-\alpha}}e^{\tilde{Y}_{t}}\\
&  =(N_{0,t}^{0,b-\delta})^{\frac{\alpha}{\alpha-1}}(\Gamma_{0,t}^{0,\gamma
})^{\frac{1}{1-\alpha}}e^{\tilde{Y}_{t}}\big[-g(t,\tilde{Z}_{t})+\frac{1}%
{2}\parallel\tilde{Z}_{t}\parallel^{2}+\frac{1}{2}\frac{\alpha}{(1-\alpha
)^{2}}\parallel b_{t}+\gamma_{t}-\delta_{t}\parallel^{2}+\frac{1}{1-\alpha
}\tilde{Z}_{t}^{\prime}\gamma_{t}\\
&  \ \ \ +\frac{\alpha}{1-\alpha}\tilde{Z}_{t}^{\prime}(b_{t}-\delta
_{t})\big]dt+(N_{0,t}^{0,b-\delta})^{\frac{\alpha}{\alpha-1}}(\Gamma
_{0,t}^{0,\gamma})^{\frac{1}{1-\alpha}}e^{\tilde{Y}_{t}}\big[\frac{1}%
{1-\alpha}\gamma_{t}^{\prime}+\frac{\alpha}{1-\alpha}(b_{t}^{\prime}%
-\delta_{t})+\tilde{Z}_{t}^{\prime}\big]dW_{t}.
\end{align*}
By the martingale principle, $\forall\delta\in\mathcal{B}_{2}^{\prime
},\ \forall\gamma\in\mathcal{B}_{2}$,
\[
-g(t,\tilde{Z}_{t})+\frac{1}{2}\parallel\tilde{Z}_{t}\parallel^{2}+\frac{1}%
{2}\frac{\alpha}{(1-\alpha)^{2}}\parallel b_{t}+\gamma_{t}-\delta_{t}%
\parallel^{2}+\frac{1}{1-\alpha}\tilde{Z}_{t}^{\prime}\gamma_{t}+\frac{\alpha
}{1-\alpha}\tilde{Z}_{t}^{\prime}(b_{t}-\delta_{t})\geq0.
\]

Define
\[
g(t,z)=\inf\limits_{\delta\in\mathcal{B}_{2}^{\prime},\gamma\in\mathcal{B}%
_{2}}\big[\frac{1}{2}\frac{\alpha}{(1-\alpha)^{2}}\parallel b_{t}+\gamma
_{t}-\delta_{t}\parallel^{2}+\frac{1}{1-\alpha}z^{\prime}\gamma_{t}%
+\frac{\alpha}{1-\alpha}(b_{t}-\delta_{t})^{\prime}z+\frac{1}{2}\parallel
z\parallel^{2}\big].
\]
Since $b_{t}$ is a deterministic function, we know that $g(t,z)$ is
deterministic and the solution $\tilde{Z}$ of (\ref{BSDEqua2}) equals $0$.
Note that
\[
g(t,0)=\frac{1}{2}\frac{\alpha}{(1-\alpha)^{2}}\inf\limits_{\delta
\in\mathcal{B}_{2}^{\prime},\gamma\in\mathcal{B}_{2}}\parallel b_{t}%
+\gamma_{t}-\delta_{t}\parallel^{2}.
\]

Let $(\hat{\delta},\hat{\gamma})$ be any continuous functions which attain the
minimum of $g(t,0)$. Then, they also attain the infimum in problem
(\ref{tildeV2}).

We have
\[
\tilde{V}(\zeta)=\frac{1-\alpha}{\alpha}\zeta^{\frac{\alpha}{\alpha-1}%
}E\big[(N_{0,T}^{0,b-\hat{\delta}})^{\frac{\alpha}{\alpha-1}}(\Gamma
_{0,T}^{0,\hat{\gamma}})^{\frac{1}{1-\alpha}}\big],\ \zeta>0.
\]
The second infimum in (\ref{dua11}) is attained at
\[
\hat{\zeta}=x^{\alpha-1}\big(E[(N_{0,T}^{0,b-\hat{\delta}})^{\frac{\alpha
}{\alpha-1}}(\Gamma_{0,T}^{0,\hat{\gamma}})^{\frac{1}{1-\alpha}}%
]\big)^{1-\alpha}=x^{\alpha-1}e^{\frac{\alpha}{2(1-\alpha)}\int_{0}^{T}%
|b_{r}+\hat{\gamma}_{r}-\hat{\delta}_{r}|^{2}dr}.
\]
Thus, the optimal terminal wealth is given by
\begin{align*}
\hat{\xi} &  =I(\hat{\zeta}\frac{N_{0,T}^{0,b-\hat{\delta}}}{\Gamma
_{0,T}^{0,\hat{\gamma}}})\\
&  =x\big(E[(N_{0,T}^{0,b-\hat{\delta}})^{\frac{\alpha}{\alpha-1}}%
(\Gamma_{0,T}^{0,\hat{\gamma}})^{\frac{1}{1-\alpha}}]\big)^{-1}(N_{0,T}%
^{0,b-\hat{\delta}})^{\frac{1}{\alpha-1}}(\Gamma_{0,T}^{0,\hat{\gamma}%
})^{\frac{1}{1-\alpha}}\\
&  =xe^{-\frac{\alpha}{2(1-\alpha)^{2}}\int_{0}^{T}|b_{r}+\hat{\gamma}%
_{r}-\hat{\delta}_{r}|^{2}dr}(N_{0,T}^{0,b-\hat{\delta}})^{\frac{1}{\alpha-1}%
}(\Gamma_{0,T}^{0,\hat{\gamma}})^{\frac{1}{1-\alpha}}.
\end{align*}

It is easy to prove the following proposition.

\begin{proposition}
$(\hat{\delta},\hat{\gamma})$ may not be unique. But the optimal terminal
wealth is unique. Furthermore, we have that for $t\in\lbrack0,T]$ and
$i=1,...,d$,%
\[
\left\{
\begin{array}
[c]{lc}%
\hat{\delta}_{it}=\varepsilon_{i}\text{ and }\hat{\gamma}_{it}=-K_{i}, &
\text{when }b_{it}>K_{i}+\varepsilon_{i};\\
\hat{\delta}_{it}=-\varepsilon_{i}\text{ and }\hat{\gamma}_{it}=K_{i}, &
\text{when }b_{it}<K_{i}+\varepsilon_{i};\\
\hat{\delta}_{it}-\hat{\gamma}_{it}=b_{it}, & \text{when }|b_{it}|\leq
K_{i}+\varepsilon_{i}.
\end{array}
\right.
\]

\end{proposition}

Now we employ the dynamic programming principle to calculate the optimal
wealth process, the optimal portfolio strategiesas as well as the utility
intensity process.

Suppose that the wealth of an large investor is $x$ at time $t$. The wealth
equation is
\begin{equation}%
\begin{cases}
dX_{s}^{t,x}=(\pi_{s}^{\prime}b_{s}-\varepsilon^{\prime}|\pi_{s}|)ds+\pi
_{s}^{\prime}dW_{s},\ s\in\lbrack t,T],\\
X_{t}^{t,x}=x\geq0,
\end{cases}
\label{wealthder}%
\end{equation}
where $\pi\in\mathcal{\bar{A}}(x;t,T)$ (recall (\ref{admispi})). The recursive
utility is
\begin{equation}
Y_{s}^{t,x}=u(X_{T}^{t,x})-\int_{s}^{T}K^{\prime}|Z_{r}^{t,x}|dr-\int_{s}%
^{T}Z_{r}^{t,x^{\prime}}dW_{r},\;s\in\lbrack t,T].\label{BSDEder}%
\end{equation}

Then the dynamic version of our problem is%
\begin{align}
& \mathrm{Maximize}\ Y_{t}^{t,x},\label{optmder}\\
& s.t.%
\begin{cases}
\pi\in\mathcal{\bar{A}}(x;t,T),\\
(X^{t,x},\pi^{t,x})\ \ \ \mathrm{satisfies\ \ \ Eq.}(\ref{wealthder}),\\
(Y^{t,x},Z^{t,x})\ \ \ \mathrm{satisfies\ \ \ Eq.}(\ref{BSDEder}),
\end{cases}
\nonumber
\end{align}
Define the value function
\[
v(t,x)=\sup\limits_{\pi\in\mathcal{\bar{A}}(x)}Y_{t}^{t,x}.
\]
$v(t,x)$ satisfies the following HJB equation (refer to \cite{Pe1}):
\begin{equation}%
\begin{cases}
\frac{\partial v}{\partial t}+\sup\limits_{\pi\in\mathbb{R}^{d}}%
\big[\frac{\partial v}{\partial x}(\pi_{t}^{\prime}b_{t}-\varepsilon^{\prime
}|\pi_{t}|)+\frac{1}{2}\frac{\partial^{2}v}{\partial x^{2}}\pi_{t}^{\prime}%
\pi_{t}-K^{\prime}|\frac{\partial v}{\partial x}\pi_{t}|\big]=0,\\
v(T,x)=u(x)=\frac{1}{\alpha}x^{\alpha}.
\end{cases}
\label{gHJB}%
\end{equation}

For any $\gamma\in\mathcal{B}_{2}$, $(\Gamma_{0,t}^{0,\gamma})$ is a
martingale. So we can define a new probability measure $P^{\gamma}$ on
$\mathcal{F}_{T}$ via
\[
P^{\gamma}(A)=E[\Gamma_{0,T}^{0,\gamma}I_{A}],\ \forall A\in\mathcal{F}_{T}.
\]
Let $E^{\gamma}[\cdot]$ be the expectation operator with respect to
$P^{\gamma}$. Under $P^{\gamma}$, the process
\[
W_{t}^{\gamma}:=W_{t}-\int_{0}^{t}\gamma_{s}ds,\ t\in\lbrack0,T]
\]
is a Brownian motion.

Thus, by the results in section 3, Problem (\ref{optmder}) is equivalent to
the following problem%
\begin{equation}%
\begin{array}
[c]{l}%
\mathrm{Maximize}\ E[\Gamma_{t,T}^{0,\hat{\gamma}}u(X^{t,x}(T))]=E^{\hat
{\gamma}}[u(X^{t,x}(T))],\\
s.t.%
\begin{cases}
\pi\in\mathcal{\bar{A}}(x),\\
(X^{t,x},\pi^{t,x})\ \ \ \mathrm{satisfies\ the\ following\ equation}\text{
}(\ref{wealthder2}),
\end{cases}
\end{array}
\label{optmder2}%
\end{equation}

\begin{equation}
\label{wealthder2}%
\begin{cases}
dX^{t,x}_{s}=(\pi^{\prime}_{s}b_{s}-\pi^{\prime}_{s}\hat\delta_{s}%
)ds+\pi^{\prime}_{s}dW_{s}=(\pi^{\prime}_{s}b_{s}+\pi^{\prime}_{s}\hat
\gamma_{s}-\pi^{\prime}_{s}\hat\delta_{s})ds+\pi^{\prime}_{s}dW^{\hat\gamma
}_{s}\\
X^{t,x}_{t}=x\geq0.
\end{cases}
\end{equation}

Therefore, $v(t,x)$ also satisfies the following HJB equation:
\begin{equation}%
\begin{cases}
\frac{\partial v}{\partial t}+\sup\limits_{\pi\in\mathbb{R}^{d}}%
\big[\frac{\partial v}{\partial x}(\pi_{t}^{\prime}b_{t}+\pi_{t}^{\prime}%
\hat{\gamma}_{t}-\pi_{t}^{\prime}\hat{\delta}_{t})+\frac{1}{2}\frac
{\partial^{2}v}{\partial x^{2}}\pi_{t}^{\prime}\pi_{t}\big]=0,\\
v(T,x)=u(x)=\frac{1}{\alpha}x^{\alpha}.
\end{cases}
\label{cHJB}%
\end{equation}

It is easy to verify the following theorem.

\begin{theorem}
(\ref{gHJB}) and (\ref{cHJB}) share the same solution $v(t,x)=\frac{1}{\alpha
}x^{\alpha}e^{\frac{\alpha}{2(1-\alpha)}\int_{t}^{T}(b_{r}+\hat{\gamma}%
_{r}-\hat{\delta}_{t})^{2}dr}$. For $s\in\lbrack t,T]$, the optimal portfolio
strategy is
\[
\hat{\pi}_{s}=-\frac{\frac{\partial v}{\partial x}}{\frac{\partial^{2}%
v}{\partial x^{2}}}(b_{s}+\hat{\gamma}_{s}-\hat{\delta}_{s})=\frac{1}%
{1-\alpha}\hat{X}_{s}^{t,x}(b_{s}+\hat{\gamma}_{s}-\hat{\delta}_{s}),
\]
and the utility intensity process is
\begin{align*}
\hat{Z}_{s} &  =-\frac{(\frac{\partial v}{\partial x})^{2}}{\frac{\partial
^{2}v}{\partial x^{2}}}(b_{s}+\hat{\gamma}_{s}-\hat{\delta}_{s})\\
&  =\frac{1}{1-\alpha}(\hat{X}_{s}^{t,x})^{\alpha}e^{\frac{\alpha}%
{2(1-\alpha)}\int_{s}^{T}(b_{r}+\hat{\gamma}_{r}-\hat{\delta}_{t})^{2}%
dr}(b_{s}+\hat{\gamma}_{s}-\hat{\delta}_{s}),
\end{align*}
where $\hat{X}_{s}^{t,x}$ is the following optimal wealth process
\begin{align*}
\hat{X}_{s}^{t,x} &  =\frac{E\big[N_{t,T}^{0,b-\hat{\delta}}I(\hat{\zeta}%
\frac{N_{t,T}^{0,b-\hat{\delta}}}{\Gamma_{t,T}^{0,\hat{\gamma}}}%
)\big|\mathcal{F}_{s}\big]}{N_{t,s}^{0,b-\hat{\delta}}}\\
&  =\frac{x}{e^{\frac{\alpha}{2(1-\alpha)^{2}}\int_{t}^{T}(b_{r}+\hat{\gamma
}_{r}-\hat{\delta}_{t})^{2}dr}}\frac{E[(N_{t,T}^{0,b-\hat{\delta}}%
)^{\frac{\alpha}{\alpha-1}}(\Gamma_{t,T}^{0,\hat{\gamma}})^{\frac{1}{1-\alpha
}}|\mathcal{F}_{s}]}{N_{t,s}^{0,b-\hat{\delta}}}\\
&  =\frac{x}{e^{\frac{\alpha}{2(1-\alpha)^{2}}\int_{t}^{s}(b_{r}+\hat{\gamma
}_{r}-\hat{\delta}_{t})^{2}dr}}\Big(\frac{\Gamma_{t,s}^{0,\hat{\gamma}}%
}{N_{t,s}^{0,b-\hat{\delta}}}\Big)^{\frac{1}{1-\alpha}}.
\end{align*}

\end{theorem}

By this theorem, we know the large investor will invest $\frac{1}{1-\alpha
}(b_{s}+\hat{\gamma}_{s}-\hat{\delta}_{s})$ percent of his wealth to the
stocks. Especially, when $|b_{it}|\leq K_{i}+\varepsilon_{i}$, the investor
will not invest on the $i$th stock at all.

\begin{remark}
Suppose that all the coefficients are deterministic and $(\bar{\theta}_{t})$,
$(\underline{\theta}_{t})$ are two given $d$-dimensional processes. Suppose
that $\bar{\theta}_{t}\geq\underline{\theta}_{t},$ a.e. on $[0,T]$. Then, our
method can tackle with the following wealth equation
\begin{equation}%
\begin{cases}
dX_{t}=(r_{t}X_{t}+\underline{\theta}_{t}^{\prime}\pi_{t}^{+}-\bar{\theta}%
_{t}^{\prime}\pi_{t}^{-})dt+\pi_{t}^{\prime}dW_{t},\\
X_{0}=x_{0},\;t\in\lbrack0,T]
\end{cases}
\end{equation}
where $\pi^{+}$ and $\pi^{-}$\ denotes the $d$-dimensional vector with $i$th
component $\pi_{i}^{+}$ and $\pi_{i}^{-},\ i=1,...,d$, respectively. This kind
of wealth equation describes the different expected returns for long and short
position of the stocks which is appeared in Jouini and Kallal \cite{JK} and El
Karoui et al \cite{EPQ1}. It also appeared in El Karoui et al \cite{EPQ} when
there are taxes which must be paid on the gains made on the stocks.
\end{remark}

\begin{remark}
When $b$ is a bounded progressively measurable processes, $g(t,z)$ is no
longer deterministic. We can prove the quadratic BSDE (\ref{BSDEqua2}) has a
unique solution $(\tilde{Y}_{t},\tilde{Z}_{t})$. There exists a pair
$(\hat{\delta},\hat{\gamma})$ attains the infimium of $g(t,\tilde{Z}_{t})$ due
to the boundness, closeness and convexity of $\mathcal{B}_{2}$ and $\mathcal{B}%
_{2}^{\prime}$.
\end{remark}

\bigskip

%%%%%%%%%%%%%%%%%%%%%%%²Î¿¼ÎÄÏ×
\renewcommand{\refname}


{\large References}

\begin{thebibliography}{99}                                                                                               %


\bibitem {BMZ}B. Bian, S. Miao, H. Zheng, \emph{Smooth value functions for a
class of nonsmooth utility maximization problems}. SIAM Journal on Financial
Mathematics, 2(2011), pp. 727-747.

\bibitem {CE}Z. Chen, L. Epstein, \emph{Ambiguity, risk, and asset returns in
continuous time}, Econometrica, 70(2002), pp. 1403-1443.

\bibitem {CC}D. Cuoco, J. Cvitanic, \emph{Optimal consumption choices for a
`large' investor}, Journal of Economic Dynamics and Control, 22(1998), pp. 401-436.

\bibitem {CK}J. Cvitanic, I. Karatzas. \emph{Convex duality in constrained
portfolio optimization}, The Annals of Applied Probability, 2(1992), pp. 767-818.

\bibitem {CK2}J. Cvitanic, I. Karatzas, \emph{Generalized Neyman-Pearson lemma
via convex duality}, Bernoulli, 7(2001), pp. 79-97.

\bibitem {DE}D. Duffie, L. Epstein, \emph{Stochastic differential utility},
Econometrica, 60(1992), pp. 353-394.

\bibitem {El}N. El Karoui, \emph{Les aspects probabilistes du controle
stochastique}, Lecture Notes in Mathematics, Springer-Verlag, 1981, pp. 73-238.

\bibitem {FMM}W. Faidi, A Matoussi, M. Mnif, \emph{Maximization of recursive
utilities: A dynamic maximum principle approach}. SIAM Journal on Financial
Mathematics, 2(2011), pp. 1014-1041.

\bibitem {EPQ}N. El Karoui, S. Peng, M. Quenez, \emph{A dynamic maximum
principle for the optimization of recursive utilities under constraints}, The
Annals of Applied Probability, 11(2001): pp. 664-693.

\bibitem {EPQ1}N. El Karoui, S. Peng, M. Quenez, \emph{Backward stochastic
differential equations in finance}, Mathematical finance, 7(1997), pp. 1-71.

\bibitem {EJ1}L. Epstein, S. Ji, \emph{Ambiguous volatility and asset pricing
in continuous time}, The Review of Financial Study, 26(2013), pp. 1740-1786.

\bibitem {EJ}L. Epstein, S. Ji, \emph{Ambiguous volatility, possibility and
utility in continuous time}, Journal of Mathematical Economics, 50(2014), pp. 269-282.

\bibitem {HIM}Y. Hu, P. Imkeller, M. Muller, \emph{Utility maximization in
incomplete markets}[J]. The Annals of Applied Probability, 15(2005), pp. 1691-1712.

\bibitem {ji-peng}S. Ji, S. Peng, \emph{Terminal perturbation method for the
backward approach to continuous-time mean-variance portfolio selection},
Stochastic Processes and their Applications, 118(2008), pp. 952-967.

\bibitem {ji-zhou}S. Ji, X. Zhou, \emph{A maximum principle for stochastic
optimal control with terminal state constraints, and its applications,}
Commun. Inf. and Syst. 6(2006) (a special issue dedicated to Tyrone Duncan on
the occasion of his 65th birthday), pp. 321-337.

\bibitem {ji-zhou-1}S. Ji and X. Zhou, \emph{A generalized Neyman-Pearson
lemma for g-probabilities}, Probability Theory and Related Fields,
148(2010),pp. 645-669.

\bibitem {JZ}H. Jin, X. Zhou, \emph{Continuous-Time Portfolio Selection under
Ambiguity}. Mathematical Control and Related Fields, 5(2015), pp. 475-488.

\bibitem {JK}E. Jouini, H. Kallal, \emph{Arbitrage in Securities Markets with
Short-Sales Constraints}. Mathematical Finance, 5(1995), pp. 197-232.

\bibitem {KLS}I. Karatzas, J. Lehoczky, S. Shreve, \emph{Optimal portfolio and
consumption decisions for a ``small investor''\ on a finite horizon}, SIAM
Journal on Control and Optimization, 25(1987), pp. 1557-1586.

\bibitem {KLSX}I. Karatzas, J. Lehoczky, S. Shreve, G. Xu, \emph{Martingale
and duality methods for utility maximization in an incomplete market}, SIAM
Journal on Control and optimization, 29(1991), pp. 702-730.

\bibitem {KS}I. Karatzas, S. Shreve, \emph{Methods of mathematical finance}.
Springer Science and Business Media, 1998.

\bibitem {Ko}M. Kobylanski, \emph{Backward stochastic differential equations
and partial differential equations with quadratic growth}, Annals of
Probability, 28(2000), pp. 558-602.

\bibitem {MX}A. Matoussi, H. Xing, \emph{Convex duality for stochastic
differential utility}, arXiv:1601.03562, 2016.

\bibitem {PP}E. Pardoux, S. Peng, \emph{Adapted solution of a backward
stochastic differential equation}, Systems and Control Letters, 14(1990), pp. 55-61.

\bibitem {Pe1}S. Peng, \emph{A generalized dynamic programming principle and
Hamilton-Jacobi-Bellman equation}, Stochastics: An International Journal of
Probability and Stochastic Processes, 38(1992), pp. 119-134.

\bibitem {Pe}S. Peng, \emph{Backward stochastic differential equations and
applications to optimal control}, Applied Mathematics and Optimization,
27(1993), pp. 125-144.

\bibitem {Qu}M. Quenez, \emph{Optimal portfolio in a multiple-priors model},
Seminar on Stochastic Analysis, Random Fields and Applications IV. Birkhauser
Basel,(2004), PP. 291-321.

\bibitem {Sc}A. Schied, \emph{Optimal investments for robust utility
functionals in complete market models}, Mathematics of Operations Research,
30(2005), pp. 750-764.

\bibitem {WZ1}N. Westray, H. Zheng, \emph{Constrained nonsmooth utility
maximization without quadratic inf convolution}. Stochastic Processes and
their Applications, 119(2009), pp. 1561-1579.

\bibitem {WZ2}N. Westray, H. Zheng, \emph{Minimal sufficient conditions for a
primal optimizer in nonsmooth utility maximization}. Finance and Stochastics,
15(2011), pp. 501-512.
\end{thebibliography}
\end{document}